\documentclass[reprint, preprintnumbers, aps, prd, floatfix, superscriptaddress]{revtex4-2}

\usepackage{amssymb, amsmath, physics, graphicx, xcolor, hyperref}

\usepackage{rotating} 

\usepackage[normalem]{ulem}

\newcommand\sYlm[3]{\ensuremath{{}_{#1}Y^{#2 #3}}}

\newcommand{\IoA}{Institute of Astronomy, University of Cambridge, Madingley Road, Cambridge, CB3 0HA, UK}
\newcommand{\KICC}{Kavli Institute for Cosmology, University of Cambridge, Madingley Road, Cambridge, CB3 0HA, UK}
\newcommand{\DAMTP}{Department of Applied Mathematics and Theoretical Physics, Centre for Mathematical Sciences, University of Cambridge, Wilberforce Road, Cambridge, CB3 0WA, UK}

\begin{document}

\title{Black-Hole Cartography}

\author{Richard Dyer}
\email{richard.dyer@ast.cam.ac.uk}
\affiliation{\IoA}

\author{Christopher J.\ Moore}
\email{christopher.moore@ast.cam.ac.uk}
\affiliation{\IoA} \affiliation{\DAMTP} \affiliation{\KICC}

\date{\today}

\begin{abstract}
    Quasinormal modes (QNMs) are usually characterized by their time dependence; oscillations at specific frequencies predicted by black hole (BH) perturbation theory.
    QNMs are routinely identified in the ringdown of numerical relativity waveforms, are widely used in waveform modeling, and underpin key tests of general relativity and of the nature of compact objects; a program sometimes called BH spectroscopy.
    Perturbation theory also predicts a specific spatial shape for each QNM perturbation. For the Kerr metric, these are the ($s=-2$) spheroidal harmonics. 
    Spatial information can be extracted from numerical relativity by fitting a feature with known time dependence to all of the spherical harmonic modes, allowing the shape of the feature to be reconstructed; a program initiated here and that we call \emph{BH cartography}. 
    Accurate spatial reconstruction requires fitting to many spherical harmonics and is demonstrated using highly accurate Cauchy-characteristic numerical relativity waveforms.
    The loudest QNMs are mapped, and their reconstructed shapes are found to match the spheroidal harmonic predictions.
    The cartographic procedure is also applied to the quadratic QNMs -- nonlinear features in the ringdown -- and their reconstructed shapes are compared with expectations based on second-order perturbation theory. 
    BH cartography allows us to determine the viewing angles that maximize the amplitude of the quadratic QNMs, an important guide for future searches, and is expected to lead to an improved understanding of nonlinearities in BH ringdown. 
\end{abstract}

\maketitle

\section{Introduction} \label{sec:intro}

The ground-based gravitational wave (GW) observatories LIGO \cite{2015CQGra..32g4001L} and Virgo \cite{2015CQGra..32b4001A} have already observed the merger of dozens of high-mass binary black holes (BHs) \cite{2023PhRvX..13d1039A} and the current observing run is expected to yield many more observations \cite{2018LRR....21....3A, postO3graphic}.
The final stage of the merger process, known as the ringdown, is associated with the remnant object settling down into its final state. 
The ringdown can be described by perturbation theory (PT), which predicts the existence of damped sinusoidal oscillations, known as quasinormal modes (QNMs) \cite{Berti:2009kk}. 

QNMs can be used directly for parameter estimation, especially for high-mass systems where only the late inspiral is observed \cite{2006PhRvD..73f4030B, 2020PhRvD.101h4053B}, or indirectly when incorporated as part of inspiral-merger-ringdown waveform models \cite{2020PhRvD.102f4002G, 2007PhRvD..75l4018B, 2018PhRvD..98h4028C}. 
However, the most important application of QNMs is in tests of general relativity (GR), fundamental physics, and the nature of extremely compact objects.
QNMs have been used to test Hawking's area theorem and the consistency of the inspiral and ringdown signals \cite{2018PhRvD..97l4069C, Isi:2020tac}. 
The no-hair theorem states that BHs should be completely characterized by their mass and spin (charge is excluded for astrophysical BHs).
When multiple QNMs are detected, consistency between their frequencies allows us to test this in a program called BH spectroscopy \cite{Carter:1971, Dreyer:2003bv, Carullo:2018sfu}. 
This requires the clear identification of at least two QNMs and this has been the subject of much discussion in the literature \cite{2019PhRvL.123k1102I, 2022PhRvL.129k1102C, 2022PhRvD.106d3005F, 2023PhRvL.131v1402C, 2024PhRvD.110d1501C}. 
However, to date, all observations are consistent with the standard picture of a vacuum Kerr BH in GR \cite{2021arXiv211206861T}.

BH ringdown can also be studied using numerical relativity (NR). 
This is typically done by fitting QNMs to individual spherical harmonic modes of the waveform.
Because the number of QNMs present is unknown \emph{a priori}, care must be taken to avoid overfitting \cite{Mitman:2022qdl, Cheung:2022rbm}.
A related issue is the choice of the start time of the ringdown, $t_0$. 
An early start time is preferable because it maximizes the amount of signal in the ringdown. 
However, a later start time is needed for the QNM description to be valid.
The optimal choice of $t_0$ is not well defined \cite{Finch:2021iip, Bhagwat:2017tkm} and is further complicated by the possible inclusion of overtones which are known to fit the signal from as early as the time of peak strain \cite{2019PhRvX...9d1060G, 2021PhRvD.103j4048D}. 
This suggests a surprisingly early start time for the ringdown and may be the result of nonlinearities becoming trapped behind the common horizon of the merging BHs \cite{Okounkova:2020vwu}. 

Recently, there has been interest in going beyond linear PT to second order in the metric perturbation, either to allow ringdown analyses to start at earlier times or to reduce the systematic errors from the nonlinearities remaining at late times. 
A key prediction of second-order PT is the existence of quadratic QNMs (QQNMs) \cite{Lagos:2022otp, Loutrel:2020wbw, Ripley:2020xby}. 
These are sourced from a pair of linear `parent' modes, whose frequencies sum to give the QQNM frequency. 
It has been shown that not only are QQNMs potentially detectable, they can be comparable in amplitude to QNMs. 
In particular, the QQNM sourced from the square of the fundamental $\ell = m = 2$ QNM has an amplitude that is comparable to the fundamental $\ell = m = 4$ QNM. 
Including nonlinear effects will be key to improving the modeling of the ringdown \cite{Mitman:2022qdl, Cheung:2022rbm}. 

\begin{figure*}[t]
    \centering
    \includegraphics[width=0.9\textwidth]{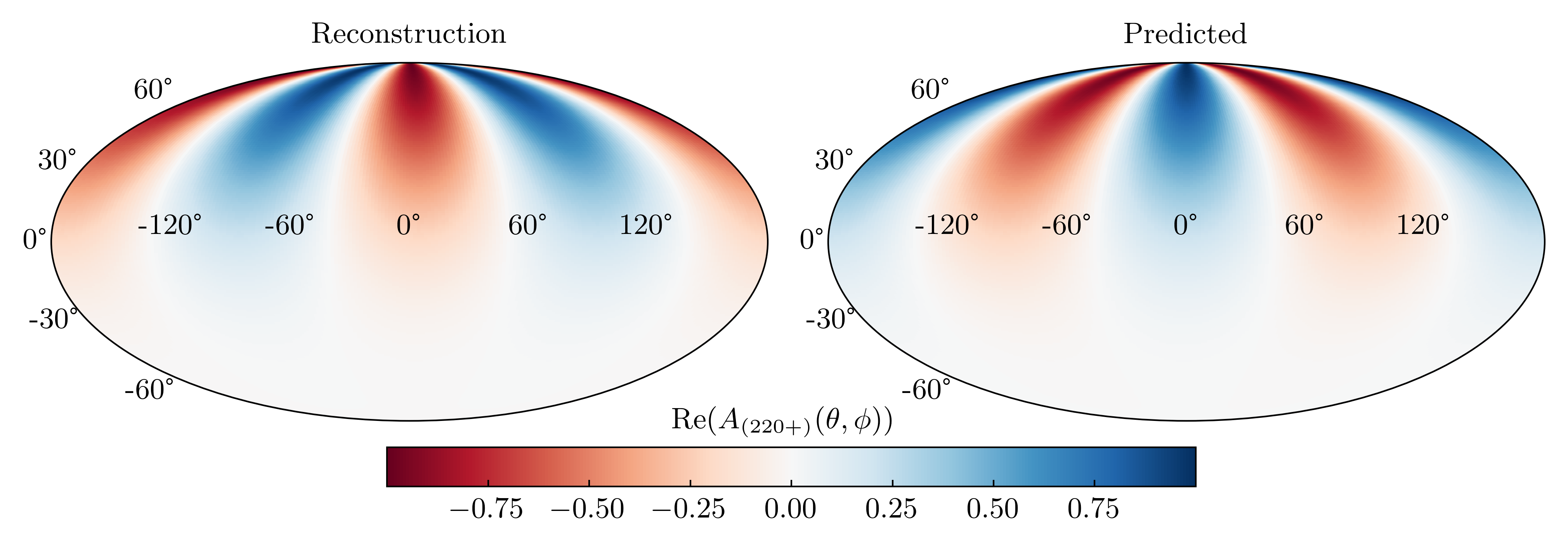}
    \caption{
        \emph{Left:} From NR, the spatial reconstruction of the fundamental QNM ($\ell=m=2$, $n=0$, prograde), normalized to agree with the corresponding spheroidal harmonic, plotted on the sphere in Mollweide projection.
        This mapping was performed for the 0001 Cauchy-characteristic extraction (CCE) waveform with a ringdown starting $t_0=8M$ after the peak of the $\ell=m=2$ mode strain (see Table \ref{tab:CCEsims}). 
        \emph{Right:} From PT, the predicted spheroidal harmonic shape.
        In both panels, the real parts of the functions are plotted and they match well up to a rotation around the $z$-axis. The spatial mismatch between these two functions is $\mathcal{M}_{\alpha}= 1.16 \times10^{-9}$ (see Sec.~\ref{subsec:mapping}).
        If the reconstruction is performed at different $t_0$, then the reconstruction is observed to rotate around the sphere with angular frequency equal to the real part of the QNM frequency.
        Because 0001 is nonprecessing, there is a symmetry between the $\pm m$ modes, and the $(\ell, m, n)=(2,-2,0)$ prograde QNM has a similar structure concentrated in the southern hemisphere.
    }
    \label{fig:linear_mapping_reconstruction}
\end{figure*}

Another motivation for studying QQNMs is their possible relationship to the GW memory effect. 
Sometimes called non-linear displacement memory, this is the permanent displacement in test particle separation after the passage of a transient GW signal \cite{Zeldovich:1974gvh, Braginsky:1985vlg, Braginsky:1987, Christodoulou:1991, Thorne:1992}. 
GW memory has not yet been observed, but is expected to be prominent in the azimuthally symmetric $m=0$ modes of the signal \cite{Pollney:2010hs, Mitman:2020pbt}. 
Pairs of prograde modes with $\pm m$ indices have frequencies that are symmetrically placed in the complex plane and hence source a QQNM with a purely imaginary frequency. These nonoscillatory modes resemble a low-frequency, persistent displacement. 
Moreover, the combination of parent modes with $\pm m$ indices gives rise to QQNMs that are azimuthally symmetric with $m=0$. 
Together with the inherent nonlinear nature of QQNMs, this hints at a connection that might allow the GW memory effect to be studied using QNMs. 

To date, most fits (using either QNMs or QQNMs) have been performed to individual spherical harmonics of the waveform.
Alternatively, the GW signal from a particular viewing direction can be fit.
These approaches do not use all of the information in the NR simulation. 
Recently multimode, or angle-averaged, fits have been used, fitting a QNM model to all spherical modes simultaneously \cite{2021arXiv211015922M}. 
By using all the modes (effectively using the signal at all viewing directions) this exploits all of the information available in the NR simulation.
Angle-averaged fits required another ingredient from PT: the predicted distribution of the GW radiation in different directions from the source.
This distribution will be called the `spatial shape' of the QNM and is a function on the 2-sphere.  
For the Kerr metric, the QNM shapes are given by the spin-weighted spheroidal harmonics \cite{Teukolsky:1973}. 
This information is usually provided in the mode-mixing coefficients that relate the spheroidal and spherical harmonics.

It is also possible to extract the QNM shapes directly from the NR data, without knowledge of the spheroidal harmonics.
By fitting a feature with common time dependence to all spherical modes and treating the mode-mixing coefficients as a free parameters, the shape can be reconstructed numerically. 
In analogy to the identification of modes using their time dependence in BH spectroscopy, this approach is called \emph{BH cartography}. 

This process can produce a \emph{spatial reconstruction} of the shape of any QNM.
An example of such a reconstruction is shown in Fig.~\ref{fig:linear_mapping_reconstruction} for the loudest fundamental QNM.
Because this reconstruction is performed using the GW signal, it is formally defined on a sphere at future null infinity. However, it is expected that this will provide insight into the structure of the perturbations that source the GW signal and are generally considered to be located deep in the strong gravitational field, near the photon sphere of the remnant BH.

BH cartography is a novel method for investigating the ringdown.
Applied to the linear QNMs, it merely allows us to verify a well-understood prediction of first-order PT, namely the spheroidal harmonics.
However, second-order perturbation theory, although capable of making unambiguous predictions for the metric perturbation, does not make such a clear and unique prediction for the QQNM shapes. There is no simple analog of the spheroidal harmonics at second order. This situation has led to different guesses, or predictions, being used for the QQNM mode shapes in the literature \cite{Lagos:2022otp, Redondo-Yuste:2023seq}.
Reconstructing the shapes of the QQNMs numerically allows us to test the different predictions.
These reconstructions also allow us to determine the viewing directions that maximize the amplitude of the QQNMs with implications for their observability. Hence, knowing the quadratic mixing will be important for future waveform modeling.

The layout of this paper is as follows. Sec.~\ref{sec:methods} describes the methods used, including a brief introduction to the key elements from BH PT (Sec.~\ref{subsec:perturbation_theory}), the least-squares fitting procedures to the NR data (Sec.~\ref{subsec:least_squares}), the new techniques for spatially mapping components of the ringdown signal (Sec.~\ref{subsec:mapping}), and the Cauchy-characteristic extraction (CCE) waveforms used in this study (Sec.~\ref{subsec:CCE}).
In Sec.~\ref{sec:linear_QNM} the mapping methods are demonstrated by applying them to the linear QNMs, which are already well understood from linear PT. This allows us to validate the BH cartography method.
Then, Sec.~\ref{sec:quadratic_QNM} applies the mapping to the QQNMs, revealing the spatial structure of the nonlinear perturbations.
Concluding remarks are presented in Sec.~\ref{sec:discussion}. 
Throughout, natural units are used in which $G=c=1$.

\section{Methods} \label{sec:methods}

Throughout, $\mathfrak{h}$ is used to denote the NR waveform and $h$ to denote a QNM model for the GW signal. 

\subsection{Ingredients from perturbation theory} \label{subsec:perturbation_theory}

GW signals contain two polarizations, and these can be conveniently combined into a single complex quantity called simply the GW strain, $h=h_+-ih_\times$.
The radially scaled GW strain can be expanded in spin-weight
$s=-2$ spherical harmonics, \sYlm{-2}{\ell}{m}, 
\begin{align}\label{eq:Ylm_expansion}
    r\mathfrak{h}(t,\theta,\phi) = M \sum_{\ell = 2}^{\infty} \, \sum_{m = -\ell}^{\ell} \mathfrak{h}^{\ell m}(t) ~ \sYlm{-2}{\ell}{m}(\theta, \phi) ,
\end{align}
where $M$ is a mass scale, usually taken to be the total irreducible mass of the source.
We adopt the convention that spherical harmonic indices are placed in superscript and, when they are introduced, spheroidal harmonic indices will be placed in the subscript.
The strain is defined at future null infinity, parameterized by the angles $\theta$ and $\phi$ on the sphere and (retarded) time $t$.
The harmonic modes $\mathfrak{h}^{\ell m}(t)$ are obtained as an output of NR simulations.

In the ringdown, linear PT suggests that the GW strain can be modeled as a sum of QNMs,
\begin{align}\label{eq:QNMmodel}
    rh^{(1)}&(t,\theta,\phi)  = 
    M \sum_{\ell, m} \sum_{n=0}^{\infty} \\ 
    \Big[ &C_{\ell m n} e^{-i \omega_{\ell m n}(t-t_0)} {}_{-2}S_{\ell m}(i\chi\omega_{\ell m n}, \theta, \phi) \nonumber \\  
    + &C'_{\ell m n} e^{-i \omega'_{\ell m n}(t-t_0)} {}_{-2}S'_{\ell m}(i\chi\omega'_{\ell m n}, \theta, \phi) \Big] , \nonumber
\end{align}
where it has been assumed, for simplicity, that the frame $(\theta,\phi)$ is adapted to the remnant BH. 
That is, the spin of the remnant BH points along the $z$-axis of the chosen coordinate system. It is always possible to perform a rotation into such a frame.
The superscript (1) in the notation is intended as a reminder that this part of the model for $h$ comes from linear PT.
The ringdown is assumed to be the perturbation of a Kerr BH, with dimensionless spin $\chi$, which is naturally described using the spheroidal harmonics, ${}_{-2}S_{\ell m}(i\chi\omega_{\ell m n}, \theta, \phi)$.

Each term in the expansion in Eq.~\ref{eq:QNMmodel} is a QNM.
The indices $\ell\geq 2$ and $|m|\leq \ell$ label angular modes while the index $n\geq 0$ labels the radial-like overtones.
For each triple $(\ell, m, n)$, there exists a pair of QNMs with different signs of the real part of their QNM frequency, $\omega'_{\ell m n} =-\omega^*_{\ell -m n}$.
The modes in this pair are sometimes referred to as the regular mode and mirror mode (denoted with a prime).
The mirror modes are associated with the modified spheroidal harmonics, ${}_{-2}S'_{\ell m}(i\chi\omega'_{\ell m n}, \theta, \phi)={}_{-2}S^*_{\ell -m}(i\chi\omega_{\ell -m n}, \pi - \theta, \phi)$.
In Eq.~\ref{eq:QNMmodel}, this pair of regular and mirror modes has been written explicitly however, it will be convenient to distinguish them with another index $p\in\{+,-\}$ for the regular and mirror modes respectively. 
Therefore, a QNM is uniquely identified by the tuple of four indices $(\ell, m, n, p)$.

The distinction between regular and mirror modes is related to the direction in which the perturbation propagates around the Kerr BH.
A mode is said to be prograde (i.e.\ co-rotating with the BH) if the signs of $m$ and $p$ are the same (if $mp>0$). 
If $mp<0$, then the mode is said to be retrograde (i.e.\ counter-rotating with the BH). 
Modes with $m=0$, along with all QNMs on Schwarzschild BHs, do not fit either the prograde or retrograde label.
Typically, in a quasicircular merger, the prograde modes are excited with significantly larger amplitudes than the retrograde modes.

The QNM model in Eq.~\ref{eq:QNMmodel} involves many indices. 
In a moment, when QQNMs are introduced, the number of indices will increase further.
Therefore, it is helpful to introduce the following abbreviated notation for the QNM model, and in particular for the spheroidal harmonics, in Eq.~\ref{eq:QNMmodel};
\begin{align}\label{eq:QNMmodel_2}
    rh^{(1)}(t,\theta,\phi) = M\sum_\alpha C_\alpha e^{-i \omega_\alpha(t-t_0)} ~ {}_{-2}S_\alpha(\theta,\phi) ,
\end{align}
where $\alpha$ denotes the tuple of QNM indices $(\ell, m, n, p)$.

The spheroidal harmonics are the natural basis for describing perturbations of Kerr BHs.
The spheroidal basis does not align perfectly with the spherical basis.
This leads to a phenomenon called mode mixing.
The mode-mixing coefficients are functions of the BH spin and are defined as
\begin{align}\label{eq:mode_mixing}
    \mu^{\ell m}_{\alpha} = \int\dd\Omega\; {}_{-2}S_{\alpha}(\theta, \phi) \big(\sYlm{-2}{\ell}{m}(\theta, \phi)\big)^*  , 
\end{align}
where $\dd \Omega$ denotes integration over the sphere parametrized by the angles $\theta$ and $\phi$.
The mode-mixing coefficients and frequencies were calculated using the \texttt{qnm} package \cite{Stein:2019mop}.

Going beyond linear PT introduces additional terms into the model in Eq.~\ref{eq:QNMmodel_2}. Notable among these are the QQNMs.
This second order part of the model, denoted with a superscript (2), is given by
\begin{align}\label{eq:QQNMmodel}
    r h^{(2)}(t,\theta,\phi) = M \sum_{\alpha} \, \sum_{\alpha'}  &C_{\alpha\alpha'} e^{-i (\omega_\alpha+\omega_{\alpha'})(t-t_0)}  \\ & \times F_{\alpha\alpha'}(\theta,\phi) . \nonumber
\end{align}
Each pair of QNMs, $(\alpha,\alpha')$, in the linear model (Eq.~\ref{eq:QNMmodel_2}) generates a QQNM in the quadratic model (Eq.~\ref{eq:QQNMmodel}).

The frequency of the QQNM is given by the sum of the frequencies of the two linear QNMs, $\omega_\alpha+\omega_{\alpha'}$.
The spatial structure of the QQNM, $F_{\alpha\alpha'}(\theta,\phi)$, is expected to be a product of two spheroidal harmonics.
There are several possibilities for how this might be done. These are discussed in Sec.~\ref{sec:quadratic_QNM} where they are compared with the numerical results.

\subsection{Least-squares fitting} \label{subsec:least_squares}

The free parameters of the QNM model are the QNM complex amplitudes $C_\alpha$.
The spheroidal harmonics and the QNM frequencies are functions of the final spin magnitude, $\chi_f$, of the remnant BH (the frequencies are also scaled by the remnant mass, $M_f$).
The quantities $\chi_f$ and $M_f$ were obtained from the NR simulation data. Specifically,
these were determined from the asymptotic Bondi data objects obtained using the \texttt{scri} package \cite{mike_boyle_2020_4041972, Boyle2013, BoyleEtAl:2014, Boyle2015a} with the relevant files from the CCE data, and are given in Table \ref{tab:CCEsims}. 

To perform model fits to the data, we use the least-squares fit implemented in the \texttt{qnmfits} package \cite{qnmfits}. 
More details on the single- and multi-mode fitting procedures are given in Appendix \ref{app:fitting}. 
For all simulations, the time $t=0$ is set to be the peak of the amplitude of the 22 mode, $|\mathfrak{h}^{22}(t)|$. 

For the NR data used in the fits, if the sums are truncated such that only modes with $\ell\leq\ell_{\rm max}$ are included, the resultant number of spherical harmonics \sYlm{-2}{\ell}{m} is $(\ell_{\rm max}+3)(\ell_{\rm max}-1)$.
For the QNM models, if the sums are truncated such that all modes with $\ell\leq\ell_{\rm max}$ and $n\leq n_{\rm max}$ are included, the number of QNMs (including both $p=\pm 1$) is $2(\ell_{\rm max}+3)(\ell_{\rm max}-1)(n_{\rm max}+1)$.

Unless otherwise stated, all of the results in this paper were obtained using $\ell_{\rm max}=8$ and $n_{\rm max}=7$. Our method requires these maximum values to be large, but the precise values are unimportant. 
For overtones with $n\geq 8$, it becomes necessary to account for the QNM multiplet connected to the presence of the algebraically special QNM (see, for example, \cite{Berti_2003}). 
The \texttt{qnm} package by default caches mode-mixing coefficients up to the value $\ell_{\rm max}=8$.
These were the practical reasons for the specific  maximum values chosen.

The modeling approach described above including all the QNMs (up to a given $\ell_{\rm max}$ and $n_{\rm max}$), regardless of whether they are expected to be significantly excited, could be described as a ``kitchen sink'' method, i.e.\ throwing everything into the model. The obvious benefit of this approach is that no important modes can be accidentally left out of the model. The potential downside is the possibility of overfitting by the large number of free parameters associated with the QNMs with small amplitudes. This is not a problem for the cartographic reconstructions performed here. This is evidenced in later results by the stability of amplitudes over a wide range of start times and by their insensitivity to the precise choices of $\ell_{\rm max}$ and $n_{\rm max}$ (see, for example, the figure and discussion in Appendix \ref{app:vary_n_max}).

For the QQNM double sum in Eq.~\ref{eq:QQNMmodel}, only individual terms were included, corresponding to the specific QQNMs under investigation. 

To measure the quality of a QNM (or QQNM) model fit to a single spherical mode of the NR data, the mismatch is used. The mode mismatch is defined as
\begin{align}\label{eq:mode_mismatch}
    \mathcal{M}^{\ell m} = 1 - \frac{|{}\braket{h^{\ell m}}{\mathfrak{h}^{\ell m}}{}|}{\sqrt{\braket{h^{\ell m}}{h^{\ell m}} \braket{\mathfrak{h}^{\ell m}}{\mathfrak{h}^{\ell m}}}},
\end{align}
where the usual signal inner product is denoted with angle brackets and is defined as
\begin{align} \label{eq:inner_prod}
    \braket{h^{\ell m}}{\mathfrak{h}^{\ell m}} = \int_{t_0}^T \dd t \; h^{\ell m}(t) \big(\mathfrak{h}^{\ell m}(t)\big)^* .
\end{align}
The modulus, denoted $|\cdot|$, in the numerator of Eq.~\ref{eq:mode_mismatch} maximizes the inner product between the model and NR data with respect to a phase shift.

An \emph{angle-averaged} mismatch is used to measure the quality of a model fit to all spherical modes. This is defined in a similar way to the mode mismatch in Eq.~\ref{eq:mode_mismatch}, but it involves all of the spherical harmonic modes and is averaged over all possible viewing directions.
The angle-averaged mismatch is defined as,
\begin{align}\label{eq:angle_averaged_mismatch}
    \mathcal{M}_{\Omega} = 1 - \frac{| {} \left[h|\mathfrak{h}\right] {} |}{\sqrt{\left[h|h\right] \left[\mathfrak{h}|\mathfrak{h}\right]}},
\end{align}
where the angle-averaged inner product, is denoted with square brackets and is defined as
\begin{align} \label{eq:angle_average_inner_prod}
    \left[h|\mathfrak{h}\right] &= \int\dd \Omega \, \int_{t_0}^T \dd t \; h(t,\theta,\phi) \mathfrak{h}(t,\theta,\phi)^* \\
    &= \sum_{\ell, m} \braket{h^{\ell m}}{\mathfrak{h}^{\ell m}}. \label{eq:angle_average_inner_prod_2}
\end{align}

\subsection{Mapping} \label{subsec:mapping}

The spatial mapping technique turns the mode-mixing coefficients from a fixed set of numbers imposed on the QNM model to a set of parameters that can be freely varied by the least-squares fitting algorithm. 

To implement this, the QNM model in Eq.~\ref{eq:QNMmodel} can be modified as follows,
\begin{align}\label{eq:mode_mapping_model}
    rh^{(1)}(t,\theta,\phi) = &\sum_{\alpha\neq\Tilde{\alpha}}\Big[C_{\alpha}e^{-i\omega_{\alpha}(t-t_0)} ~ {}_{-2}S_\alpha(\theta,\phi) \Big]  \\
    + &\sum_{\ell,m}\Big[A_{\Tilde{\alpha}}^{\ell m} e^{-i\omega_{\Tilde{\alpha}}(t-t_0)} ~ \sYlm{-2}{\ell}{m}(\theta,\phi) \Big] . \nonumber
\end{align}
Here, $\Tilde{\alpha}$ is the mode that is being spatially mapped. 
The first summation is the same as in the usual model in Eq.~\ref{eq:QNMmodel_2} and includes all modes except the one that is mapped.
The second summation has only one QNM, the mapped mode $\Tilde{\alpha}$, but includes an independent copy of it separately in each spherical harmonic.
In this model, $C_{\alpha}$ and $A^{\ell m}_{\Tilde{\alpha}}$ are regarded as free complex amplitudes to be determined from the NR data by least-squares fitting.
The mapping model in Eq.~\ref{eq:mode_mapping_model} can be expanded in spherical harmonics using the mode-mixing coefficients in Eq.~\ref{eq:mode_mixing},
\begin{align} \label{eq:mode_mapping_model_2}
    & r h^{(1)}(t,\theta,\phi) = \sum_{\ell,m} r h^{(1)}{}^{\ell m}(t) \sYlm{-2}{\ell m}{}(\theta, \phi),\quad \mathrm{where} \\
    & r h^{(1)}{}^{\ell m}(t) = A^{\ell m}_{\Tilde{\alpha}}e^{-i\omega_{\Tilde{\alpha}}(t-t_0)} + \sum_{\alpha\neq\Tilde{\alpha}}C_{\alpha}\mu^{\ell m}_\alpha e^{-i\omega_{\alpha}(t-t_0)} \nonumber .
\end{align}
In this form, it is more easily seen how the new free amplitude parameters $A^{\ell m}_{\Tilde{\alpha}}$ take on the role of the mixing coefficients for the mapped mode $\Tilde{\alpha}$. 

The mapped QNM has one such amplitude for each of the $(\ell_{\rm max}+3)(\ell_{\rm max}-1)$ different $\mathfrak{h}_{\ell m}$ NR modes used in the fit (for $\ell_{\rm max}=8$, this equals 77). When all amplitudes are taken together, they contain the information required to reconstruct the spatial structure of the QNM. 

In contrast to an amplitude $C_{\alpha}$ that is obtained with an assumed spatial structure predicted from PT, the set of independent amplitudes, $A^{\ell m}_{\Tilde{\alpha}}$, does not assume any particular spatial structure. Therefore, spatial mapping can determine the spatial structure of QNMs from NR. 
Agreement between the prediction and the reconstruction gives another way to verify the presence of a QNM in the data.
The shape of the reconstructed mode is given by,
\begin{align}\label{eq:Athetaphi}
    A_{\Tilde{\alpha}}(\theta,\phi) = \sum_{\ell, m} A_{\Tilde{\alpha}}^{\ell m} \sYlm{-2}{\ell}{m}(\theta,\phi).
\end{align}
This can be visualized directly, e.g.\ by plotting on the sphere $\theta,\phi$ using any cartographic projection (see, for example, the Mollweide projection plots in Figs.~\ref{fig:linear_mapping_reconstruction}, \ref{fig:reconstruction_t0}, \ref{fig:q_fundamental_mapping_reconstruction} and \ref{fig:reconstruction_memory}).
Alternatively, when assessing how accurately the spatial structure of a mode has been recovered compared to a theoretical prediction, we use the recovered $A^{\ell m}_{\Tilde{\alpha}}$ coefficients to define yet another mismatch,
\begin{align} \label{eq:spatial_mismatch}
    \mathcal{M}_{\Tilde{\alpha}} = 1-\frac{|{}\{A_{\Tilde{\alpha}}|{}_{-2}S_{\Tilde{\alpha}}\}{}|}{\sqrt{ 
    \{A_{\Tilde{\alpha}}|A_{\Tilde{\alpha}}\} \{{}_{-2}S_{\Tilde{\alpha}}|{}_{-2}S_{\Tilde{\alpha}}\} 
    }},
\end{align}
where 
\begin{align} 
    \{A_{\Tilde{\alpha}}|{}_{-2}S_{\Tilde{\alpha}}\} &= \int\dd\Omega\; A_{\Tilde{\alpha}}(\theta,\phi) \big({}_{-2}S_{\Tilde{\alpha}}(\theta,\phi)\big)^* \\
    &= \sum_{\ell, m} A_{\Tilde{\alpha}}^{\ell m} \left(\mu^{\ell m}_{\Tilde{\alpha}}\right)^* .
\end{align}
This measures the agreement between the recovered shape and the predicted spheroidal harmonic.
The modulus in the numerator of Eq.~\ref{eq:spatial_mismatch} maximizes the inner product between the reconstruction and the NR data with respect to a rotation about the $z$-axis. 

Note that the mismatch can be improved, and the overall mapping sped up, by limiting the data and model to only include modes with the same $m$ as the mapped mode.

It is also possible to extend this approach to map more than one QNM at the same time by removing another term, say $\Tilde{\alpha}'$, from the first sum in Eq.~\ref{eq:mode_mapping_model} and including another sum over $\ell,m$ for the new mapped mode $\Tilde{\alpha}'$ with coefficients $A^{\ell m}_{\Tilde{\alpha}'}$. 
However, the quality of the reconstruction (e.g.\ as quantified using $\mathcal{M}_{\Tilde{\alpha}}$) decreases as the number of mapped modes is increased. An example of this is shown in Appendix \ref{app:multiQNMmapping}. 
The details of the changes to the least-squares fitting procedure needed for the spatial mapping are detailed in Appendix \ref{app:fitting}.

\subsection{CCE waveforms} \label{subsec:CCE}

\begin{table}
\caption{\label{tab:CCEsims} 
    CCE waveforms used in this study.
    All NR data were obtained from the public catalog \cite{SXS_CCE_catalog}.
    The final column gives the best ringdown start time that minimizes the angle-averaged mismatch $\mathcal{M}_\Omega$ for a QNM model with $\ell_{\rm max}=8$ and $n_{\rm max}=7$.
    }
\begin{ruledtabular}
\begin{tabular}{r|llccr}
  ID & $q$ & Descriptive name & $M_f\, [M]$ & $\chi_f$ & $t_0\, [M]$ \\
  \hline
  0001 & 1 & q1\_nospin & 0.95162 & 0.6864 & 17 \\
  0002 & 1 & q1\_aligned\_chi0\_2 & 0.94551 & 0.7464 & 21 \\
  0003 & 1 & q1\_aligned\_chi0\_4 & 0.93759 & 0.8038 & 23 \\
  0004 & 1 & q1\_aligned\_chi0\_6 & 0.92684 & 0.8578 & 26 \\
  0005 & 1 & q1\_antialigned\_chi0\_2 & 0.95159 & 0.6863 & 17 \\
  0006 & 1 & q1\_antialigned\_chi0\_4 & 0.95149 & 0.6859 & 17 \\
  0007 & 1 & q1\_antialigned\_chi0\_6 & 0.95133 & 0.6851 & 17 \\
  0008 & 1 & q1\_precessing & 0.95758 & 0.6033 & 11 \\
  0009 & 1 & q1\_superkick & 0.94943 & 0.6786 & 29 \\
  0010 & 4 & q4\_nospin & 0.97793 & 0.4716 & 16 \\
  0011 & 4 & q4\_aligned\_chi0\_4 & 0.97026 & 0.6860 & 12 \\
  0012 & 4 & q4\_antialigned\_chi0\_4 & 0.97200 & 0.6720 & 17 \\
  0013 & 4 & q4\_precessing & 0.98089 & 0.4298 & 6 \\
\end{tabular}
\end{ruledtabular}
\end{table}

\begin{figure*}[t]
\begin{minipage}[b]{0.48\linewidth}
\centering
\includegraphics[width=\textwidth]{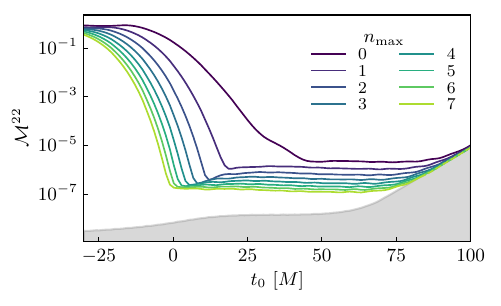}
\end{minipage}
\hspace{0.5cm}
\begin{minipage}[b]{0.48\linewidth}
\centering
\includegraphics[width=\textwidth]{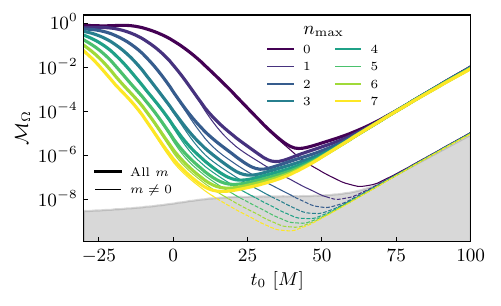}
\end{minipage}
\caption{
    Mismatches for linear QNM overtone model fits to the NR simulation 0001. 
    In both panels, the gray region indicates the NR error estimate described in Appendix \ref{app:nr_error}. When a curve first dips below the numerical error, the line style changes to dashed.
    \emph{Left:} The individual mode mismatch in the $\ell=m=2$ spherical harmonic mode. 
    The fit uses a QNM model with fundamental plus overtones, including the $\alpha=(2,2,n,+)$ modes with $n\in\{0,1,\ldots,n_{\rm max}\}$. 
    \emph{Right:} The angle-averaged mismatch calculated across all (thick lines) $\ell,m$ spherical harmonic modes up to $\ell_{\rm max}=8$.
    The model includes all QNMs up to those with $\ell_{\rm max}=8$ and $n=n_{\rm max}$.
    As with the single-mode case, increasing the number of overtones reduces the overall mismatch and shifts the minima of the mismatch to earlier times, associated with an earlier starting time for the ringdown, but it does so more smoothly in the angle-averaged case. 
    A significantly improved angle-averaged match is achieved if the axisymmetric $m=0$ spherical harmonics are excluded from this fit (thin lines). These modes are poorly fit by QNMs.
}
\label{fig:M22_Momega}
\end{figure*}

\begin{figure*}[t]
\begin{minipage}[b]{0.48\linewidth}
\centering
\includegraphics[width=\textwidth]{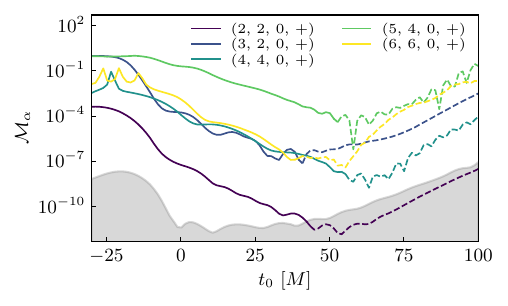}
\end{minipage}
\hspace{0.5cm}
\begin{minipage}[b]{0.48\linewidth}
\centering
\includegraphics[width=\textwidth]{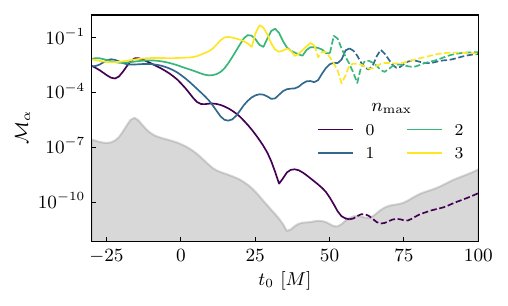}
\end{minipage}
\caption{
    The spatial mismatch between spatially reconstructed QNMs and the expected spheroidal harmonics.
    \emph{Left:} The spatial mismatch for the loudest QNM for each $\ell$ in simulation 0001 up to $\ell = 6$. Because $q=1$, the loudest mode in the $\ell=3$ sector is the $m=2$ mode similar for the other odd $\ell$ values. The grey region indicates the NR error in the (2,2,0,+) mode mapping, described in Appendix \ref{app:nr_error}.
    The other modes have different (larger) associated errors which are not shown here, but the change in the style of each from solid to dashed indicates the point where the spatial mismatches first drop below their corresponding error thresholds.
    \emph{Right:} The spatial mismatch for the overtones $\Tilde{\alpha}=(2,2,n_{\rm max},+)$ in simulation 0001.
    The gray region indicates the NR error in the $n=0$ QNM, and the solid to dashed transition in the other lines indicates the point where the spatial mismatch first drops below its corresponding error threshold. The first three overtones are shown and their reconstructions get progressively worse but show minima in the spatial mismatches at earlier start times, consistent with their faster decay rates.
    The higher-order overtones have worse spatial mismatches and are not shown.
}
\label{fig:Malpha}
\end{figure*}

This work uses NR waveforms from the Spectral Einstein Code (SpEC), found in the Simulating Extreme Spacetimes catalog \cite{Boyle:2019kee, SXS_CCE_catalog}.
Analysis was performed with the 13 publicly available simulations with waveforms obtained using CCE. These waveform data were created using the SpECTRE CCE module \cite{spectrecode, Moxon:2020gha, Moxon2023}.  

Currently, only a limited number of CCE waveforms are available in the public catalog (see Table \ref{tab:CCEsims}). 
These simulations sparsely cover a range of progenitor parameters, such as spin magnitudes, orientations, and the binary mass ratio, $q$. 
All simulations were transformed into the superrest frame at $300M$ after the peak strain using the \texttt{scri} package. 
The waveforms were subsequently shifted in time so that the peak of the $22$ mode amplitude, $|\mathfrak{h}^{22}(t)|$, occurs at $t=0$ \cite{Mitman:2024uss, Mitman:2022kwt, Mitman:2021xkq, 2021arXiv211015922M}. 

The CCE method of waveform extraction is more faithful to the true GW strain than waveforms obtained by extrapolation \cite{Moxon:2020gha}.
This improvement is particularly important for studies of the ringdown using multiple QNMs.
This is also true of the mapping procedure developed here.

For each simulation taken from the catalog, the highest resolution level and the second smallest worldtube radius were together used as the preferred CCE waveform.
Waveforms from other levels and worldtube radii were used to estimate the size of the NR errors, as described in Appendix \ref{app:nr_error}.

\section{Linear QNM results} \label{sec:linear_QNM}

The new spatial mapping methods are demonstrated by first applying them to linear QNMs.
In this case, first-order PT makes a clear, unambiguous prediction for the shape of the QNMs -- the spin-weight $s=-2$ spheroidal harmonics. 
This provides a way to test the BH cartography program.

Figure ~\ref{fig:M22_Momega} shows the mismatch results for simple QNM fits to simulation 0001.
The left panel shows the mismatch in the $\ell=m=2$ spherical mode when fitted using a QNM model with a variable number of overtones. This is similar to Fig.~1 in Ref \cite{2019PhRvX...9d1060G}.
The right panel shows the angle-averaged mismatch calculated using all modes when fitted using a QNM model including all the QNMs (up to $\ell_{\rm max}$ and $n_{\rm max}$). 

Both panels in Fig.~\ref{fig:M22_Momega} display a similar structure, starting with a rapidly declining mismatch approaching a minimum at a time on or after $t_0 = 0$.
At later times, the mismatch increases as the signal decays. 
The minima get pushed to earlier start times with an increasing number of overtones, related to the possibility of using an earlier choice of start time. 
The start times that give the minimum angle-averaged mismatch for the $n_{\mathrm{max}} = 7$ QNM model are listed for each simulation in Table \ref{tab:CCEsims}; these start times are used in subsequent figures.

The accuracy of the angle-averaged QNM fits are limited by the $m=0$ spherical harmonic modes.
In the right panel of Fig.~\ref{fig:M22_Momega}, the thin lines indicate the angle-averaged mismatch with $m=0$ modes excluded from the fit. 
There is a considerable improvement in the mismatches, down to the level expected given the numerical noise present in the simulation.
This shows that the $m=0$ are fit less well by the QNM model than all the other $m$ modes.
This may be due to the presence of the nonoscillatory GW memory signal in these modes. There may also be an imperfect mapping to the superrest frame for these modes, affecting fits at late times. 

Having established that the QNM models are able to faithfully represent the NR data, we now turn to mapping.
Figure \ref{fig:Malpha} shows the results of spatially mapping various QNMs. 
The left panel shows the spatial mismatch of some of the loudest fundamental ($n=0$) QNMs in simulation 0001. 
Because this is a simulation of equal-mass quasicircular binary BHs ($q=1$), the symmetry $\phi\rightarrow\phi+\pi$ forces all spherical harmonic modes with odd $m$ indices to be identically zero. Therefore, the loudest mode in, for example, the $\ell=3$ sector is the $(\ell,m)=(3,2)$ mode. 
The right panel shows the spatial mismatch for some of the $\ell=m=2$ QNM overtones, with the largest overtone in the model being mapped in each case, analogously to the right panel in Fig.~\ref{fig:M22_Momega}. 
In both panels, the mismatch of the fundamental mode is considerably smaller than that of the subdominant modes. Generally, the mismatch values decrease with the increasing amplitude of the QNM. 

In the figures above, the mismatches are plotted as a function of the start time.
If the mode reconstruction is performed with a different choice of start time, then the values obtained for the $A^{\ell m}_{\Tilde{\alpha}}$ amplitudes are different.
Typically, for a reasonable range of $t_0$ values in the ringdown, the magnitude of the amplitude decreases exponentially, and the phases of the amplitudes increase linearly with increasing $t_0$.
The decreasing amplitude is due to the exponential decay of the QNM with a timescale given by the reciprocal of the imaginary part of $\omega_{\Tilde{\alpha}}$.
The changing phase corresponds to the rotation of the perturbation sourcing the QNM around the symmetry axis of the BH.
The rotation occurs at a frequency related to the real part of $\omega_{\Tilde{\alpha}}$. To track this rotation, the following angle can be used,
\begin{align}
    \Phi_\alpha = \mathrm{arg} \big(\{ A_{\Tilde{\alpha}} \big| {}_{-2}S_{\Tilde{\alpha}} \} \big).
\end{align} 

\begin{figure}[t]
    \centering
    \includegraphics[width=0.49\textwidth]{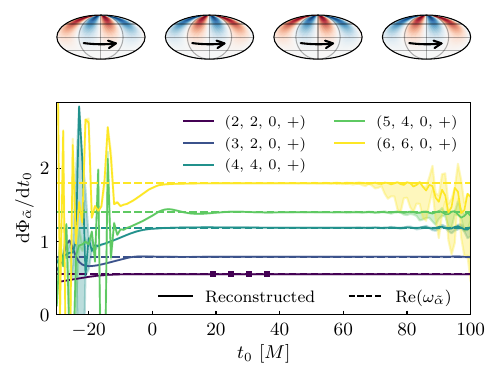}
    \caption{The top four figures show the real part of the spatial reconstruction of the (2,2,0,+) QNM at the start times indicated by squares on the main plot. These demonstrate the rotation of the QNM around the sphere. 
    \emph{Main plot:} Derivative of the argument $\Phi_\alpha$ of the spatial mismatch (corresponding to the phase shift between the mapped and predicted spatial structures) with respect to ringdown start time $t_0$.
    The solid lines show this quantity for simulation 0001 for the same 5 QNMs (matching colors) used in the left panel of Fig.~\ref{fig:Malpha}. The shaded regions give an estimate of the error in the reconstructed frequencies and are described in Appendix \ref{app:nr_error}.  
    The horizontal dashed lines give the predicted rotation frequencies, which are the real part of the complex QNM frequencies, $\omega_{\Tilde{\alpha}}$. An animated version of these outcomes showing the rotation around the BH is available at \cite{Spatial_mapping_repo}.
    }
    \label{fig:reconstruction_t0}
\end{figure}

The rate of change of this angle with the ringdown start time $t_0$ is plotted in Fig.~\ref{fig:reconstruction_t0} for the same QNMs used in the left panel of Fig.~\ref{fig:Malpha}. 
The values agree with the predicted real part of the QNM frequencies. 
In terms of the spatial reconstruction, this corresponds to a rotation of the QNM around the sphere indicated by the top four panels taken at the start times indicated by squares on the lower panel. The spatial mapping technique uses the QNM frequency to determine the spatial structure, so this is not a demonstration of inferring the frequency from a free parameter. Instead, the figure shows that the frequency can be reconstructed from the phase of the values of $A^{\ell m}_{\Tilde{\alpha}}$ over a range of start times.

\begin{figure}[t]
    \centering
    \includegraphics[width=0.48\textwidth]{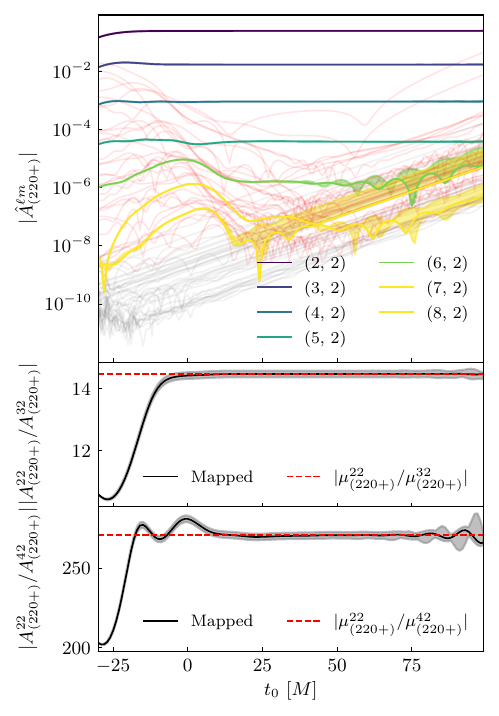}
    \caption{
    The magnitudes of the decay-corrected amplitudes, $|\hat{A}^{\ell m}_{(220+)}|$, obtained from spatial mapping of the (2,2,0,+) mode for the 0001 waveform are plotted. The thick colored lines show the spherical modes indicated in the legend that are expected to have large amplitudes. Translucent grey lines indicate odd $m$ modes, and the translucent red lines indicate even $m \neq 2$ modes. The ratios of selected amplitudes are plotted in the lower panels. The values for the ratios of the spheroidal harmonics predicted by linear PT (expressed in terms of ratios of mode-mixing coefficients) are plotted as horizontal dashed lines. The translucent area surrounding the lines on the three panels indicates the NR uncertainty in the values obtained. The way this is determined is described in Appendix \ref{app:nr_error}. 
    }
    \label{fig:amplitude_ratios_t0}
\end{figure}

To demonstrate the mapping to each spherical mode, the top panel in Fig.~\ref{fig:amplitude_ratios_t0} shows the individual mode amplitudes in a reconstruction of the fundamental $\Tilde{\alpha}=(2,2,0,+)$ QNM.
For clarity, the decay-corrected amplitudes are plotted; that is each mode amplitude is plotted multiplied by an exponential factor to correct for its expected decay rate so that $|\hat{A}^{\ell m}_{(220+)}| = |A^{\ell m}_{(220+)}e^{i\omega_{(220+)}(t - t_0)}|$. 

The bold colored lines are the $m=2$ modes given in the legend. A consistent amplitude, across a range of start times, in the $\ell \leq 4$ modes indicates that this QNM is being confidently detected.
It is necessary that a QNM is detected in at least two spherical harmonics in order for there to be a sensible mapping reconstruction.
In this representation of the results, the familiar mode-mixing coefficients are related to the vertical separation of the horizontal lines on this plot.

The $m \neq 2$ modes have also been included as translucent lines. In a multimode fit, all mixing coefficients where $m \neq 2$ would be zero and, therefore, so would the corresponding amplitudes. 
In contrast, here, where all spherical modes have been included in the spatial mapping, these amplitudes appear as nonzero values. The translucent gray lines moving diagonally along the figure are vanishing spherical modes, which are being scaled by the decay correction absorbed into the amplitude. 
The remaining translucent red lines are spherical modes for which the spatial mapping has attributed some nonzero amplitude. 
The contrast in the structure of these modes and the $m=2$ modes suggests that spatial mapping is correctly identifying the relevant spatial structure of the $(2,2,0,+)$ mode. The nonzero amplitudes of the $m \neq 2$ modes can be considered an indication of the numerical accuracy of the spatial mapping technique. 

The bottom panels in Fig.~\ref{fig:amplitude_ratios_t0} show the magnitude of the ratios of the amplitudes in the $(2,2)$ and either $(3,2)$ or $(4,2)$ spherical modes respectively. These values agree with the magnitudes of the ratios of the corresponding mixing coefficients, which are also plotted. This provides an alternative mode-specific way to verify the accuracy of the spatial mapping technique beyond using the spatial mismatch. 
This is particularly helpful in checking that the reconstruction is not simply a consequence of overfitting to the loudest spherical mode (which, in this case, would be $\ell = m = 2$).

\begin{figure}[t]
    \centering
    \includegraphics[width=0.49\textwidth]{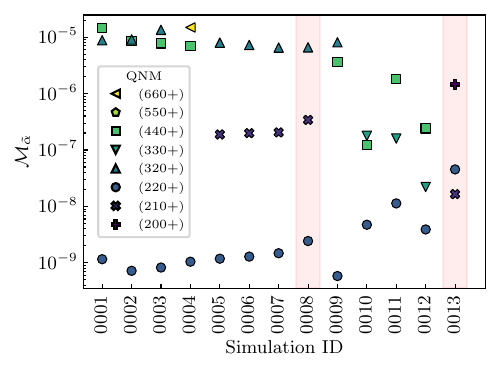}
    \caption{
        The spatial mismatch for the best three reconstructed modes for all CCE waveforms. Highlighted columns indicate precessing simulations. In all cases, it was possible to reconstruct at least three QNMs with a spatial mismatch of $\lesssim 10^{-5}$ with their associated spheroidal harmonics. In all cases, except the precessing and asymmetric 0013 simulation, the best-mapped mode is the fundamental $(2,2,0,+)$ QNM. The identity of the next best-mapped QNMs changes depending on the parameters of the progenitor binary. In all cases, the best three mapped QNMs are all prograde $n=0$ modes.
    }
    \label{fig:catalog_plot}
\end{figure}

Finally, Fig.~\ref{fig:catalog_plot} gives the QNMs with the lowest mismatch obtained when spatially mapping every mode individually, for each simulation. 
As noted in Fig. \ref{fig:M22_Momega} and again here, the $(2,2,0,+)$ mode -- which is typically loudest -- is almost always reconstructed considerably more accurately than subdominant modes. The exception is simulation 0013, which is an asymmetric binary with precessing spins, resulting in the unusual combination and ordering of the best three reconstructed modes.  

The analysis in this section demonstrates that spatial mapping can effectively reconstruct the spatial structure of QNMs without explicitly providing mode-mixing coefficients. At first order, where PT makes clear predictions about the spatial structure, the technique offers another method for detecting QNMs and may, in analogy with the single- and multi-mode fits in Fig. \ref{fig:M22_Momega}, provide another way to determine the start time of the ringdown. 

\begin{figure*}[t]
    \centering
    \includegraphics[width=0.99\textwidth]{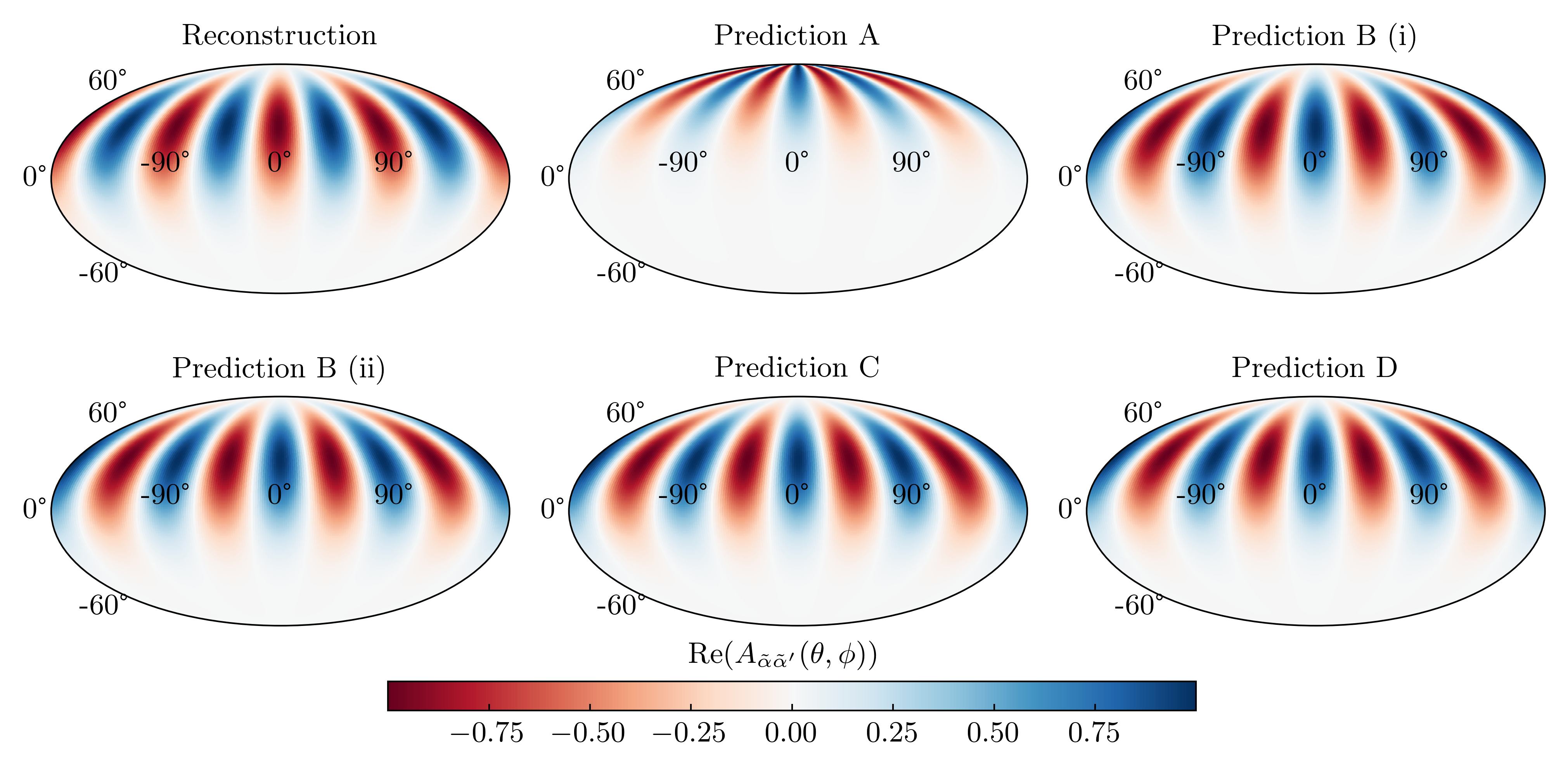}
    \caption{
        \emph{Leftmost panel:} The real part of the normalized spatial reconstruction of the dominant QQNM with $\Tilde{\alpha}=\Tilde{\alpha}'=(2,2,0,+)$ for the 0001 NR simulation. \emph{Other panels:} The various predictions for the shape of this QQNM discussed in the main text. Prediction A clearly has qualitatively the wrong shape and is, therefore, discarded. The remaining B, C, and D predictions all have qualitatively similar shapes (from B(i) to D, the spatial mismatches with the reconstruction are $\mathcal{M}_{\Tilde{\alpha}}= 7.4\times 10^{-4}$, $\mathcal{M}_{\Tilde{\alpha}} = 2.9 \times 10^{-4}$, $\mathcal{M}_{\Tilde{\alpha}} = 2.5 \times 10^{-4}$, and $\mathcal{M}_{\Tilde{\alpha}} = 2.6 \times 10^{-4}$) and are further investigated in Fig.~\ref{fig:theta_max}.
        }
    \label{fig:q_fundamental_mapping_reconstruction}
\end{figure*}

\section{Quadratic QNM results} \label{sec:quadratic_QNM}

In principle, second-order PT is capable of making an unambiguous prediction for the second-order part of the metric perturbation, including all of the QQNMs. While work in this direction has been attempted \cite{Lagos:2022otp, Ma_2024}, there is no analog of the spheroidal harmonics for QQNMs, i.e.\ there does not exist a simple, universal family of functions that describe the shape of these features for all perturbed BHs.

In the absence of complete calculations at second order, there have been a number of guesses, or predictions, made in the literature for the spatial shape of the QQNMs. These are inspired by second-order PT. Because the QQNM is sourced by two linear QNMs, some of these assume that the shape $F_{\alpha \alpha'}$ (see Eq.~\ref{eq:QQNMmodel}) will be given by a quadratic combination of spheroidal harmonics. In this section, several of these incomplete second-order PT predictions are outlined and compared to the spatially reconstructed modes.

\textbf{Prediction A:} A naive guess for these functions is to take the product of two parent spheroidal harmonics, ${}_{-2}S_{\alpha}(\theta,\phi)$ and ${}_{-2}S_{\alpha'}(\theta,\phi)$. Then, the spatial shape of the QQNM takes the form 
\begin{align} \label{eq:predA}
    \mathbf{A:}\quad F_{\alpha\alpha'}(\theta,\phi)\propto {}_{-2}S_{\alpha}(\theta,\phi) {}_{-2}S_{\alpha'}(\theta,\phi).
\end{align}

\begin{figure}[t]
    \centering
    \includegraphics[width=0.49\textwidth]{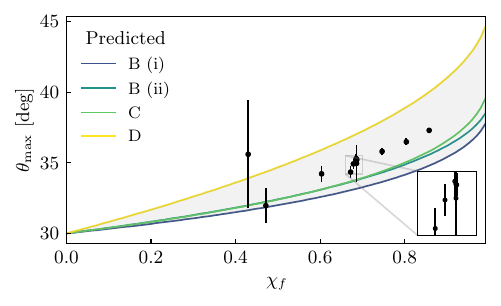}
    \caption{
        The values of $\theta_{\mathrm{max}}$ for the dominant QQNM with $\Tilde{\alpha}=\Tilde{\alpha}'=(2,2,0,+)$ are given for all simulations in Table \ref{tab:CCEsims}, plotted as a function of the dimensionless spin of the final remnant BH, $\chi_f$. For each simulation, the reconstruction was performed at the start time $t_0$ indicated in Table \ref{tab:CCEsims} to obtain the central marker and for a range of start times $\pm 5 M$ about this time to determine the uncertainty estimates shown in the error bars. The predicted values for this angle are plotted as smooth curves obtained using the B(i), B(ii), C and D predictions, with the grey shading indicating the region within which reconstructed values of $\theta_{\mathrm{max}}$ are expected to be.
    }
    \label{fig:theta_max}
\end{figure}

However, while this gives qualitatively the correct dependence on the azimuthal $\phi$ coordinate, it gives the wrong $\theta$ dependence (see, for example, the reconstruction of the dominant QQNM in Fig.~\ref{fig:q_fundamental_mapping_reconstruction}). 
In particular, it predicts that the dominant QQNM's largest observed amplitude would be at the north pole ($\theta=0$). 
This is in disagreement with the reconstruction, which shows a finite angle $\theta_{\rm max}\approx 35^{\circ}$ for this maximum. 
This is further quantified in Fig.~\ref{fig:theta_max}, where $\theta_{\mathrm{max}}$ is plotted for the $\alpha=\alpha'=(2,2,0,+)$ QQNM for all simulations. 

This prediction cannot be correct because the resulting $F_{\alpha\alpha'}$ does not have the required -2 spin weight for a GW field. Instead, $F_{\alpha \alpha'}$ in Eq.~\ref{eq:predA} has a spin weight of -4 resulting from the fact that the product of a spin-weight $s$ function and a spin weight $s'$ function has spin-weight $s+s'$.

\textbf{Prediction B:} This issue can be resolved by using a product of spheroidal harmonics with spin weights that sum to -2. 
There is not a unique choice for this.
The possibilities for the fundamental QQNM with $\alpha=\alpha'=(2,2,0,+)$ are
\begin{align}
    \mathbf{B}\,\mathbf{(i):}\quad F_{\alpha \alpha'}(\theta,\phi) &\propto {}_{-2}S_{\alpha}(\theta,\phi) \, {}_{0}S_{\alpha'}(\theta,\phi) , \label{eq:predBi} \\
    \mathbf{B}\,\mathbf{(ii):}\quad F_{\alpha \alpha'}(\theta,\phi) &\propto {}_{-1}S_{\alpha}(\theta,\phi) \, {}_{-1}S_{\alpha'}(\theta,\phi).
\end{align}
For other QQNMs with $\alpha\neq \alpha'$, there is no symmetry between the spin-weight indices in Eq.~\ref{eq:predBi} and the order in which the spheroidal harmonics are used gives additional possibilities.
Furthermore, for QQNMs that involve higher $\ell$ indices, there are more possible combinations involving spheroidal harmonics with $|s|\geq 3$. (Note, the spin weight is constrained to be $|s|\leq \ell$.)

\textbf{Prediction C:} There is another proposal given in \cite{Redondo-Yuste:2023seq}.
This assumes that the shape of the QQNM is given by a single spheroidal harmonic which, rather than being defined by its indices, has a modified spheroidicity defined using the frequency $\omega_\alpha+\omega_{\alpha'}$ of the QQNM,
\begin{align} \label{eq:predC}
    \mathbf{C:}\quad F_{\alpha \alpha'}(\theta,\phi) \propto  {}_{-2}S_{\lambda\mu}\big(i\chi_f\left[\omega_{\alpha}+\omega_{\alpha'}\right],\theta,\phi\big),
\end{align}
where $\lambda=\ell+\ell'$ and $\mu=m+m'$ and, for clarity, we have temporarily reverted to the  original notation for the spheroidal harmonics introduced in Eq.~\ref{eq:QNMmodel}.
Here, the notation for the spheroidal harmonics from Ref.~\cite{Stein:2019mop} is used.

\textbf{Prediction D:} An alternative resolution to the spin-weight issue is to take the product of two parent spheroidal harmonics with spin weights that do not sum to $-2$, but act on them with the appropriate spin-weight raising or lowering operator to obtain the required spin-weight $-2$ function \cite{Keefe}. For example, taking the product of two spin-weight $-2$ spheroidal harmonics as in prediction A but acting on them twice using the spin-weight raising operator gives,
\begin{align} \label{eq:predD}
    \mathbf{D:}\quad F_{\alpha\alpha'}(\theta,\phi)\propto \eth^2 \left( {}_{-2}S_{\alpha}(\theta,\phi) {}_{-2}S_{\alpha'}(\theta,\phi)\right).
\end{align}

For each of these predictions, a corresponding quadratic mode-mixing coefficient $\mu^{\ell m}_{\alpha \alpha'}$ can be computed. 
These three predictions are compared in Fig. \ref{fig:q_fundamental_mapping_reconstruction} against the reconstruction of the fundamental QQNM in simulation 0001.
Prediction A is clearly disfavored and is not discussed further.
Visually, there is little difference between the other predictions.

Interestingly, the B, C and D predictions all correctly suggest that the projected amplitude of the fundamental QQNM is maximized at a nonzero viewing angle, away from the axis of the remnant BH's final spin.
This has important consequences for the observability of this QQNM in GW data.
This angle is investigated further in Fig.~\ref{fig:theta_max} where it is plotted as a function of the spin magnitude of the remnant for all remaining predictions.
This angle depends only mildly on the remnant BH spin. The projected amplitude of the fundamental QQNM is always maximized by a viewing angle in a narrow range around $30^\circ \lesssim \theta_{\rm}\lesssim 45^\circ$. For a nonspinning remnant, the predictions reduce to a value of $30^\circ$, corresponding to the maximum amplitude angle of the $\sYlm{-2}{4}{4}$ spherical harmonic. As spin increases, the contribution of higher $\ell$ spherical modes becomes increasingly significant, resulting in the unique structures identified in the figure. The maximum amplitude viewing angles of the reconstructed simulations all live (within their error bars) in the narrow range given by the predictions.

\begin{figure}[t]
    \centering
    \includegraphics[width=0.49\textwidth]{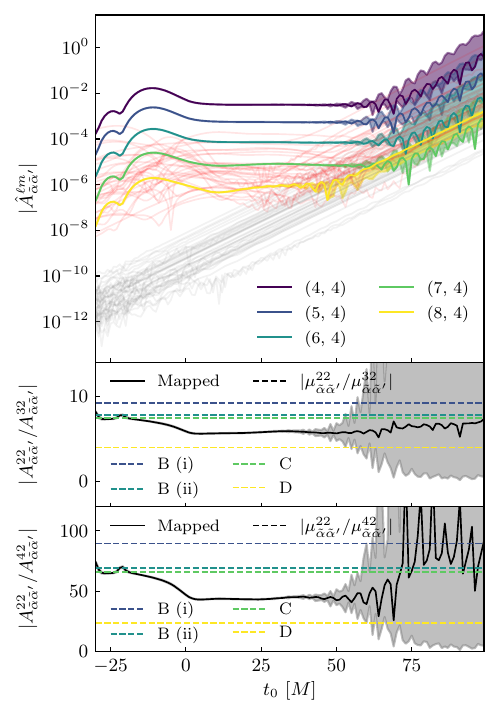}
    \caption{
    The magnitude of the decay-corrected amplitudes obtained from spatial mapping of the fundamental quadratic mode with $\Tilde{\alpha}=\Tilde{\alpha}'=(2,2,0,+)$ for simulation 0004. The solid lines indicate the $m=4$ spherical modes, whilst the translucent grey lines indicate odd $m$ modes, and the translucent red lines indicate even $m \neq 4$ modes. The ratios of selected amplitudes are plotted in the lower panels. The values for the structure predicted by second-order PT (expressed in terms of ratios of second-order mode-mixing coefficients) are plotted as horizontal dashed lines. The translucent areas surrounding the lines on the three panels indicate the NR uncertainty in the values obtained.
    }
    \label{fig:amplitude_ratios_t0_quad}
\end{figure}

In general, error bars are larger for lower remnant spins. This is most clearly demonstrated in the two simulations with the smallest spins, which are considerably harder to map, resulting in a greater sensitivity to the start time used. This effect may be due to higher spin remnants having longer ringdowns, which provides more signal on which to perform the mapping, leading to smaller errors. 

Figure \ref{fig:amplitude_ratios_t0_quad} shows the reconstructed amplitudes $A^{\ell m}_{\Tilde{\alpha} \Tilde{\alpha}'}$ for the fundamental QQNMs in simulation 0004 as a function of the start time.
These (decay-corrected) amplitudes show a clear stable behavior across a wide range of start times for several different $\ell$ indices. This is important because, as noted above, in order for these reconstructions to be meaningful, it is necessary to detect the feature in multiple spherical harmonics.

For the equal-mass simulation 0004, the reconstructions do not appear to match any prediction. 
As noted in Ref.~\cite{Lagos:2022otp}, second-order PT does not make a unique prediction for the shape of the QQNMs.
This is in contrast with linear PT which predicts a spheroidal harmonic for each QNM.
At second order, it is instead possible that the true spatial structure may be formed from some combination of the predicted modes that cannot be determined analytically.

\begin{figure}[t]
    \centering
    \includegraphics[width=0.48\textwidth]{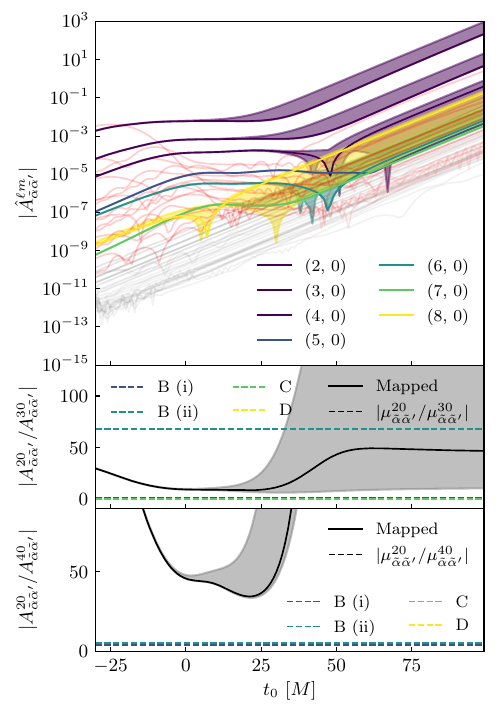}
    \caption{
    The magnitude of the decay-corrected amplitudes obtained from spatial mapping of the QQNM $\Tilde{\alpha} = (2,2,0,+)$ and $\Tilde{\alpha}' = (2,-2,0,-)$ for simulation 0004. The solid lines indicate the $m = 0$ spherical modes, whilst the translucent grey lines indicate odd $m$ modes, and the translucent red lines indicate even $m \neq 0$ modes. The ratios of selected amplitudes are plotted in the lower panels. The values for the structure predicted by second-order PT (expressed in terms
    of ratios of second-order mode-mixing coefficients) are plotted
    as horizontal dashed lines. The translucent areas surrounding
    the lines on the three panels indicate the NR uncertainty in
    the values obtained.  
    }
    \label{fig:amplitude_ratios_t0_memory}
\end{figure} 

\begin{figure}[t]
    \centering
    \includegraphics[width=0.48\textwidth]{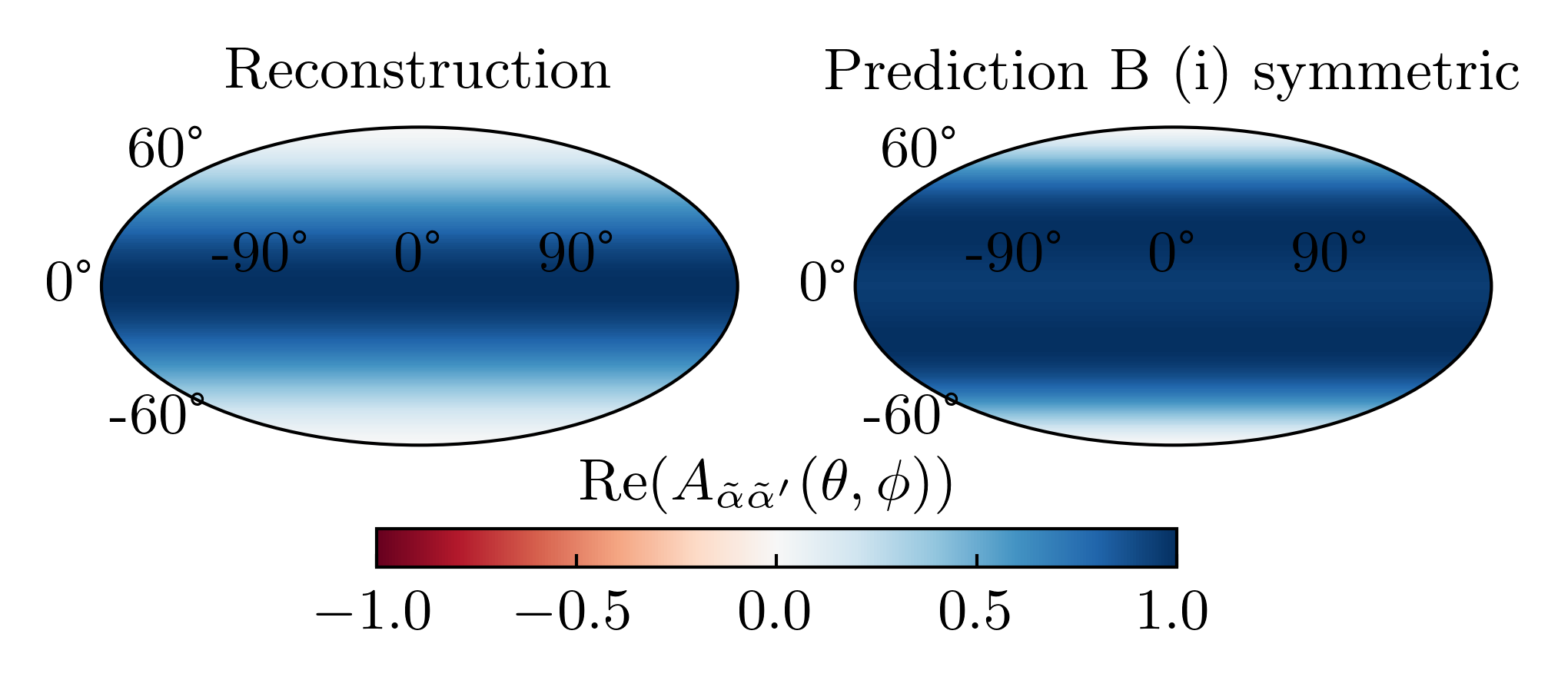}
    \caption{The real part of the spatial reconstruction of the $\Tilde{\alpha} = (2,2,0,+), \Tilde{\alpha}' = (2,-2,0,-)$ QQNM for the 0004 waveform, plotted as a function of the angles on the sphere using a Mollweide projection. The spatial structure is compared to B (i) with a symmetric combination of spin weights, which has the required axisymmetry of the QQNM along the equator, where the amplitude is expected to be loudest.
    The agreement between the spatial reconstruction and the prediction is poor, which is reflected in the mismatch of $\mathcal{M}_{\alpha} = 3.2 \times 10^{-2}$.}  
    \label{fig:reconstruction_memory}
\end{figure}

\subsection{The $m=0$ modes} \label{subsec:memory}

Among all possible QQNMs, there is a subset of modes with parent QNMs $\alpha=(\ell,m,n,p)$ and $\alpha'=(\ell,-m,n,-p)$.
These modes have purely imaginary frequencies and are, therefore, non-oscillatory, unlike most other QNMs or QQNMs. 
Because the azimuthal indices of the linear QNMs in the pair sum to zero, the resulting QQNM appears in the azimuthally symmetric $m=0$ sector.
Additionally, if $mp>0$, then because both linear QNMs in the pair are prograde and typically excited with large amplitudes, the resulting QQNM is, therefore, also expected to have a large amplitude. 
These properties suggest a connection between this subset of QQNMs and the GW memory effect.

To investigate these further, the $\Tilde{\alpha}=(2,2,0,+)$, $\Tilde{\alpha}'=(2,-2,0,-)$ QQNM is spatially mapped, and the amplitudes and their ratios are plotted in Fig. \ref{fig:amplitude_ratios_t0_memory}. In principle, this combination of prograde parent modes should be the loudest $m=0$ mode in the ringdown, if it is indeed present. The stability of the mapped amplitudes is considerably weaker than in the previous cases, which may indicate that this QQNM is too quiet to be clearly detected. However, flat regions in the amplitudes and the upper amplitude ratio give some indication that the mapping is successfully identifying the mode and its corresponding mixing. 

The mode is reconstructed in Fig. \ref{fig:reconstruction_memory}. Choosing option B (i), with an equal combination of $s=-2$, $s'=0$ and $s=0$, $s'=-2$, gives the required axisymmetric structure. However, it does not match with the overall expected shape. No individual prediction matches the spatially reconstructed mode. However, again, it is possible that the correct structure is some combination of modes.

\section{Discussion} \label{sec:discussion}

This paper has demonstrated that it is possible to determine the specific spatial shape of QNMs and QQNMs. Using CCE waveforms, it has been shown that the angular structure of QNMs can be accurately reconstructed in agreement with first-order PT. In analogy with single- and multi-mode ringdown fits, the spatial mapping procedure has a mismatch -- called the spatial mismatch -- which, correspondingly, has a minimum that represents the region around which the QNM is clearly present in the ringdown. 

The spatial mapping procedure is effective across many simulations and QNMs, including the mapping of several QNMs simultaneously. The full spatial shape can be plotted which, using the phase information contained in the mapping amplitudes, can be used to reconstruct the real part of the QNM frequency. This was found to be associated with the rotation of the modes around the sphere. 

In contrast to first order, second-order PT makes no clear prediction about the spatial structure of QQNMs. Hence, spatial mapping presents a method to obtain this structure for a given QQNM and remnant black hole. This is likely to be useful in future precision waveform modeling, where second-order effects will be important. 

The spatial mapping has been applied to what is expected to be the loudest QQNM, demonstrating that this method can determine the spatial structure for second-order PT. The technique was also applied to the loudest $m=0$ QQNM. This mode is special due to a possible connection with the memory effect. We demonstrated that it may be possible to reconstruct this mode, providing another way to investigate this connection. 

While the predictions considered in this work varied, we have shown that they were generally qualitatively very similar to each other and the reconstructed spatial structure. Perhaps the most significant finding is that the loudest $\Tilde{\alpha} = \Tilde{\alpha}' = (2,2,0,+)$ QQNM was not visible at $\theta = 0$. Instead, the projected amplitude was maximized at a viewing angle of $\theta \approx 35^{\circ}$ for a remnant with spin $\chi_f \approx 0.7$. This will hopefully provide a guide for future searches of QQNMs in GW data.


\begin{acknowledgments}
    We thank Keefe Mitman, Eliot Finch, Julian Westerweck, and Maximiliano Isi for their helpful comments on a draft of this paper. We also thank the anonymous referee for their very helpful report.
    Finally, we thank Keefe Mitman for organizing the public release of the high-fidelity CCE simulations used in this study.
\end{acknowledgments}

\clearpage


\bibliographystyle{apsrev4-1}
\bibliography{references}



\onecolumngrid

\appendix

\section{Single- and multi-mode fitting} \label{app:fitting}

This appendix describes the single- and multi-mode least-squares fitting used in this paper, carried out using the \texttt{qnmfits} package, in addition to the modifications required to perform spatial mapping. 
Here, in addition to the abbreviated notation $\alpha = (\ell, m, n, p)$ for the indices of the various QNMs, a similar abbreviated notation $\beta = (\ell, m)$ for the spherical-mode indices is also introduced. Recall, spherical $\beta$ indices are placed in superscript while spheroidal $\alpha$ indices are placed in subscript. 
The perturbation order is suppressed in the notation of this appendix.
The strain time series is sampled at discrete times $t_k$, with $k=0,1,\ldots,K-1$. 
The summation convention is not used; all summations are written explicitly. 

Starting from the QNM model in Eq.~\ref{eq:QNMmodel_2} and expanding the spheroidal harmonics in terms of the spherical harmonics using the mode-mixing coefficients in Eq.~\ref{eq:mode_mixing}, gives the following QNM model for a single spherical harmonic mode:
\begin{align}\label{eq:QNMmodelsinglemode}
    r h^{\beta}(t) &=
    M \sum_{\alpha} C_{\alpha} \mu^{\beta}_{\alpha} \exp(-i \omega_\alpha[t-t_0]).
\end{align}
Hereafter, the factors of $r$ (on the left-hand side) and $M$ (on the right-hand side) in Eq.~\ref{eq:QNMmodelsinglemode} are dropped from our notation for simplicity.
When fitting a single spherical harmonic mode, the mode-mixing coefficient can be absorbed into a redefinition of the QNM amplitude, $\Tilde{C}_\alpha = C_{\alpha} \mu^{\beta}_{\alpha}$.
The amplitudes $\Tilde{C}_{\alpha}$, are determined by minimizing the sum of the squares of the fit residuals,
\begin{align}\label{eq:leastsquares}
    R^{\beta} = \sum_k \Big(\mathfrak{h}^{\beta}(t_k) - \sum_{\alpha} \Tilde{C}_{\alpha} \exp(-i \omega_\alpha[t_k-t_0])\Big)^2.
\end{align}

The QNM models considered here, e.g.\ in Eq.~\ref{eq:QNMmodelsinglemode}, are all linear models that depend linearly on the parameters to be determined. This would not be true if, for example, the QNM frequencies or the mass and spin of the remnant were included as free parameters.
In this sense, the fitting to be performed is simple and can be reduced to a linear algebra operation, as described below. 
This is what allows us to include such a large number of QNMs in some of the angle-averaged fits used in this paper.

In order to efficiently minimize the quantity in Eq.~\ref{eq:leastsquares}, the model in Eq.~\ref{eq:QNMmodelsinglemode} is first recast as a matrix equation $\pmb{h}^{\beta} = \pmb{a}^{\beta}\cdot\pmb{C}$. 
It is then possible to rewrite the sum of the squared residuals in Eq.~\ref{eq:leastsquares} as the Euclidean 2-norm of a matrix-vector equation,
\begin{align}\label{eq:leastsquares_matrix}
    R^{\beta} = \big|\big|\pmb{\mathfrak{h}}^{\beta} - \pmb{a}^{\beta}\cdot\pmb{C} \big|\big|^2.
\end{align}
The quantity $\pmb{\mathfrak{h}}^{\beta}$ is a vector of length $K$ containing the NR data for the spherical mode $\beta$, $\pmb{C}$ is a vector of length $N$ (the number of QNMs used in the fit; i.e.\ the number of terms in the sum over $\alpha$) containing the QNM amplitudes to be determined, and $\pmb{a}^{\beta}$ is a matrix of shape $(K,N)$ containing the exponential factors from Eq.~\ref{eq:QNMmodelsinglemode}. 
Specifically, in terms of components, these are given by $(\pmb{\mathfrak{h}}^{\beta})_i = \mathfrak{h}^\beta(t_i)$ with $\beta$ fixed, $(\pmb{a}^{\beta})_{ij}=\exp(-i\omega_j(t_i-t_0))$, and $
\pmb{C}_j=\Tilde{C}_j$.
In this form, the \texttt{numpy.linalg.lstsq} function can be used to efficiently minimize the sum of the squared residuals and find the corresponding QNM amplitudes, $\pmb{C}$.

\begin{sidewaysfigure}
    \centering
    \includegraphics[width=\linewidth]{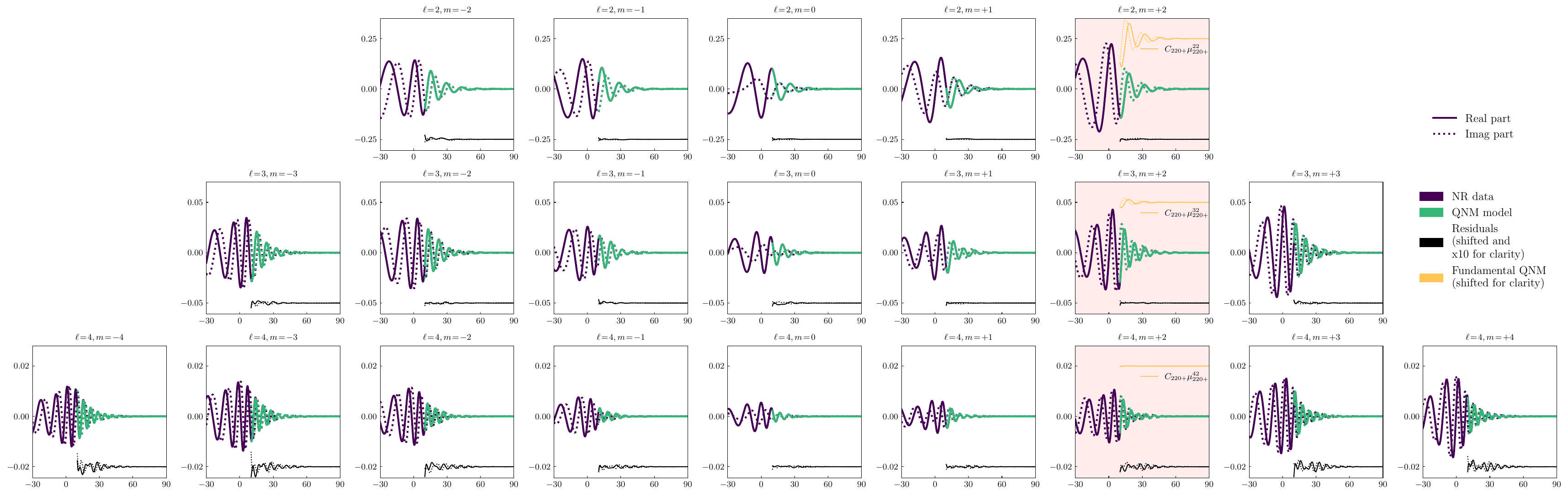}
    \caption{
        Illustration of an angle-averaged multimode QNM fit to NR data.
        Shown in blue is the NR waveform $\mathfrak{h}^{\beta}(t)$ for the CCE 0013 simulation, which, because it comes from a precessing asymmetric binary (see Table \ref{tab:CCEsims}), contains significant power spread across many spherical harmonic modes.
        Here, for the purpose of illustration, modes up to and including $\ell_{\rm max}=4$ are plotted (real [solid] and imaginary [dotted] parts shown separately), arranged in a grid with $m$ and $\ell$ increasing along the rows and down the columns, respectively.
        Note that the $y$-axis range is shared for all panels in a given row (i.e.\, with the same $\ell$) but changes down the columns, while the time range on the $x$-axis is the same in all panels.
        Shown in green is a QNM model angle-averaged fit to the NR data. 
        The QNM model, $h^{\beta}(t)$, starts at $t_0=10M$ and includes all QNMs (prograde and retrograde) up to and including those with $\ell_{\rm max}=4$ and $n_{\rm max}=1$. The total number of damped sinusoids in this fit is 84.
        The residuals (data minus model) are plotted in black on each panel, shifted vertically down, and multiplied by a factor of 10 for clarity.
        It should be emphasized that the angle-averaged fit \emph{is not} separately fitting a sum of damped sinusoids to each spherical mode; rather, the intrinsic amplitude and phase of each QNM, $C_{\alpha}$, is determined by fitting to all the spherical harmonic modes simultaneously, with its contribution in each multiplied by the mode-mixing coefficient $\mu^{\beta'}_{\alpha}$. 
        Only the amplitude $C_{\alpha}$ is free to vary in the fit -- the mode-mixing coefficients $\mu^{\beta'}_{\alpha}$ are determined by the final spin of the remnant.
        Because, as a consequence of axial symmetry, the mode-mixing coefficients satisfy $\mu^{\beta'}_{\alpha}\propto\delta_{m}^{m'}$, the mixing occurs only within the columns of this figure.
        In order to emphasize the role of mode mixing, the contribution of the fundamental QNM with $\alpha=(\ell, m, n, p)=(2,2,0,+)$ is isolated and plotted in orange (shifted vertically up for clarity) in the highlighted $m=2$ column. The formula for the amplitude of this contribution is indicated in each panel by the inset legend. 
        This fundamental QNM dominates the spherical $\ell=m=2$ mode signal, but is also significant in the $\ell=3$, $m=2$ mode, with smaller contributions for higher $\ell$ values. 
        It should be noted that the remnant BH in simulation 0013 has a small final spin of $\chi_f=0.43$, which means that the mode mixing is less prominent here than in other cases. 
    }
    \label{fig:mega_plot}
\end{sidewaysfigure}

When fitting a QNM model to multiple spherical harmonic modes simultaneously (as is done in the angle-averaged fits) it is no longer possible to ignore the mode-mixing coefficients by a redefinition of the amplitudes. 
A QNM $(\ell,m,n,p)$ generates a contribution in all the spherical modes $(\ell',m')$ with $m=m'$, and this effect is illustrated in the shaded column of plots in Fig.~\ref{fig:mega_plot}.
The QNM model in Eq.~\ref{eq:QNMmodelsinglemode} is now used for all values of $\beta$ for which NR data are available (instead of just a single value), and the QNM amplitudes $C_\alpha$ are determined by minimizing the following sum of the squares of the fit residuals summed over all spherical harmonic modes:
\begin{align}\label{eq:leastsquares_angleaveraged}
    R =  \sum_{\beta} \sum_k \Big(\mathfrak{h}^{\beta}(t_k) - \sum_{\alpha} \ C_{\alpha} \mu_{\alpha}^{\beta} \exp(-i \omega_\alpha[t_k-t_0])\Big)^2.
\end{align}
This quantity can also be written as the Euclidean 2-norm of a matrix-vector equation,
\begin{align} \label{eq:leastsquares_angleaveraged_matrix}
    R = \big|\big|\pmb{\mathfrak{h}} - \pmb{a}\cdot\pmb{C} \big|\big|^2.
\end{align}
Here, the vectors and matrices are now formed by stacking together (concatenating) all the spherical harmonic modes.
For example, the quantity $\pmb{\mathfrak{h}}$ is now a vector of length $KM$ (where $M$ is the number of spherical harmonics used in the fit, i.e.\, the number of terms in the sum over $\beta$) formed from stacking all the $K$ vectors $\pmb{\mathfrak{h}}^\beta$ together. 
The order in which the vectors are stacked is arbitrary, but must be done consistently with the construction of the matrix $\pmb{a}$.
The quantity $\pmb{a}$ is now a $(KM,N)$ matrix containing the mode-mixing and exponential factors in Eq.~\ref{eq:leastsquares_angleaveraged}. 
The quantity $\pmb{C}$ is still a vector of length $N$ containing the QNM amplitudes to be determined.
Specifically, in terms of components, these are given by $(\pmb{\mathfrak{h}})_i = \mathfrak{h}^\beta(t_k)$ with $\beta=i//K$ and $k=i\%K$, $(\pmb{a})_{ij}=\mu^{\beta}_j\exp(-i\omega_j(t_k-t_0))$ again with $\beta=i//K$ and $k=i\%K$, and $
\pmb{C}_j=C_j$. 
Here, \texttt{Python} notation is used for the modulus and integer division operations.
In this form, the \texttt{numpy.linalg.lstsq} function can again be used to efficiently minimize the sum of the squared residuals and find the corresponding QNM amplitudes, $\pmb{C}$.

An angle-averaged fit to multiple spherical harmonic modes is illustrated in Fig.~\ref{fig:mega_plot} for simulation 0013. Each panel shows both the data, $\mathfrak{h}^\beta$, and the QNM model, $h^\beta$, for one spherical harmonic mode. 
The model gives a good fit to all the modes, as can be seen by examining the residuals. 
It should be stressed that each panel in this figure is \emph{not} a separate fit -- they are all part of the same fit. 
This is necessary because of the mode-mixing effect and is illustrated graphically by highlighting the effect of a single QNM (the fundamental $\alpha=(2,2,0,+)$ mode) in different spherical harmonics (see the three plots with $m=2$ in the highlighted column).
In this example, because mode mixing only occurs between modes with the same index $m$, the different columns of this figure are, in fact, independent. 
However, when mapping (see Sec.~\ref{subsec:mapping}) the shape of a particular mode, the mode-mixing coefficients are effectively promoted to free parameters and all parts of this plot would be coupled. 

The angle-averaged fitting procedure can be extended to include a spatially mapped mode, $\Tilde{\alpha}$. Using the mapping model from Eq.~\ref{eq:mode_mapping_model} (see also Eq.~\ref{eq:mode_mapping_model_2}), the sum of the squared residuals in Eq.~\ref{eq:leastsquares_angleaveraged} now becomes
\begin{align}\label{eq:leastsquares_spatiallymapped}
    R =  \sum_{\beta} \sum_k \Big( \mathfrak{h}^{\beta}(t_k) - A^{\beta}_{\Tilde{\alpha}} \exp(-i \omega_{\Tilde{\alpha}}[t_k-t_0]) - \sum_{\alpha \neq \Tilde{\alpha}} C_{\alpha} \mu_{\alpha}^{\beta} \exp(-i \omega_\alpha[t_k-t_0]) \Big)^2 ,
\end{align}
where the parameters to be determined are now both the amplitudes $C_{\alpha}$ (for all $\alpha\neq\alpha_{\Tilde{A}}$) and $A_{\Tilde{\alpha}}^\beta$ (for all $\beta$).
This quantity can also be written as the Euclidean 2-norm of a matrix-vector equation.
In practice, this is done by starting from the matrix 
$\pmb{a}$ in the angle-averaged fit in Eq.~\ref{eq:leastsquares_angleaveraged_matrix}, identifying the column corresponding to the mode to be mapped, $j = \Tilde{\alpha}$, and replacing it with stacked matrices of shape $(K, M)$ that, in terms of components, is given by $(\pmb{\Tilde{a}})_{ij}=\delta_{\beta j}\exp(-i\omega_{\Tilde{\alpha}}(t_k-t_0))$. 
Consequently, the mapped mode is fit to each spherical mode individually, as in the single-mode case in Eq.~\ref{eq:mode_mapping_model}, absorbing the mixing coefficient into the amplitude $A^{\beta}_{\Tilde{\alpha}}$, which is now associated with a spherical mode in the model.

\section{Effect of varying the number of overtones in a QNM reconstruction} \label{app:vary_n_max}

\begin{figure*}[t]
\begin{minipage}[b]{0.32\linewidth}
\centering
\includegraphics[width=\textwidth]{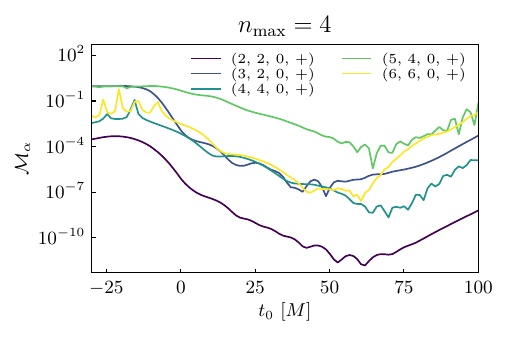}
\end{minipage}
\begin{minipage}[b]{0.32\linewidth}
\centering
\includegraphics[width=\textwidth]{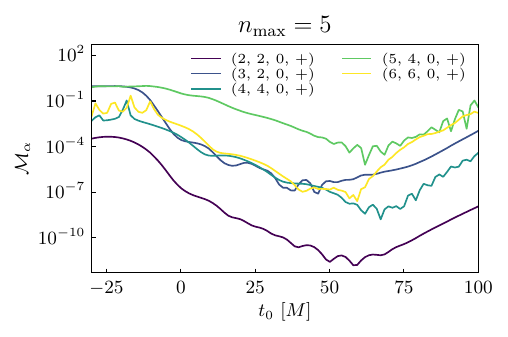}
\end{minipage}
\begin{minipage}[b]{0.32\linewidth}
\centering
\includegraphics[width=\textwidth]{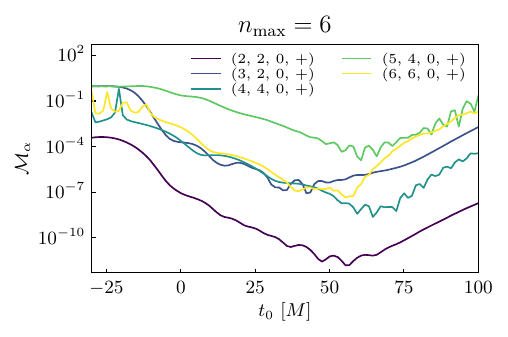}
\end{minipage}
\caption{
   The effect of changing the overtone content in spatial mapping is demonstrated by reducing $n_{\rm max}$ to 4, 5, or 6. The technique can be seen to be robust against these changes, supporting the ‘‘kitchen sink'' approach to this work. These figures should be compared with the left panel of Fig.~\ref{fig:Malpha}, which was produced using $n_{\rm max}=7$.
}
\label{fig:overtone_sensitivity}
\end{figure*}

To address the sensitivity of spatial mapping to the choice of QNM content in the model, the figures in this paper can be reproduced for varying values of $n_{\rm max}$ and $\ell_{\rm max}$. An example of this is shown in Fig.~\ref{fig:overtone_sensitivity}, where the left panel of Fig.~\ref{fig:Malpha} has been reproduced for $n_{\rm max} \in \{4, 5, 6\}$. The resulting figures are extremely similar to the $n_{\rm max} = 7$ case, highlighting the stability of the ``kitchen sink'' approach to changes in the choice of $n_{\rm max}$. 

\section{Simultaneous mapping of multiple QNMs} \label{app:multiQNMmapping}

\begin{figure}[ht]
    \centering
    \includegraphics[width=0.48\textwidth]{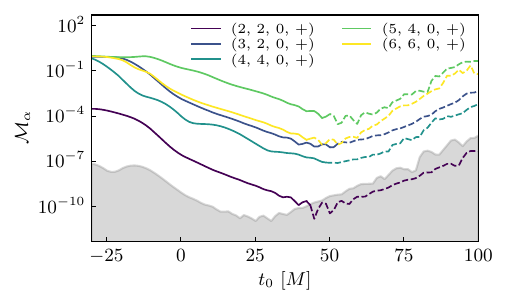}
    \caption{The spatial mismatch between spatially reconstructed QNMs and the expected spheroidal harmonics. Here, the loudest QNMs for each $\ell$ in simulation 0001 up to $\ell = 6$ are mapped simultaneously. This is in contrast with the left-hand panel of Fig.~\ref{fig:Malpha} where each QNM is spatially mapped individually. The grey region indicates the NR error in the (2,2,0,+) mode and the change in the style of each line from solid to dashed indicates the point where the spatial mismatches first drop below their corresponding error thresholds. Only QNMs and spherical modes with $m \in \{2, 4, 6\}$ are used.}
    \label{fig:sm_mismatch_loudest_multi}
\end{figure}

The spatial mapping techniques can be generalized to mapping any number of QNMs and QQNMs simultaneously. To map multiple modes, the procedure in Appendix \ref{app:fitting} is followed, replacing every column corresponding to a mapped mode with a stacked matrix. The resulting amplitude vector will then contain one value for each QNM or QQNM in the model, in addition to $M$ amplitudes for each mapped mode $\Tilde{\alpha}$. The resulting reconstructions and corresponding mismatches deteriorate with an increasing number of mapped modes, however several modes can still be mapped with remarkable accuracy. Fig. \ref{fig:sm_mismatch_loudest_multi} demonstrates the simultaneous mapping of the five loudest modes used in Fig. \ref{fig:Malpha} and Fig. \ref{fig:reconstruction_t0}, with all QNMs and spherical modes with $m\in\{2, 4, 6\}$ included in the fit. The mismatches are nonetheless comparable to Fig. \ref{fig:Malpha}, albeit slightly higher, corresponding to the loss in accuracy from removing additional spatial information from the fit. 

\section{Numerical error} \label{app:nr_error}

The numerical error, which is expected to be present in all NR simulations, is determined in several ways throughout this paper. All the methods involve comparing the preferred version of the CCE simulations (Level 5, second smallest radius) to the other possible combinations of the Levels 4 and 5, and the second and third smallest worldtube radii. 

For the plots in Fig. \ref{fig:M22_Momega}, the three other possible waveforms are each shifted in time to find the optimal alignment that minimizes the mismatch with the preferred waveform. This is done by rolling the waveform by discrete time intervals and computing the mismatch until the minimum is found, then shifting the position of $t = 0$ by this corresponding amount. After this alignment, the mismatch is calculated between the preferred simulation and the three remaining simulations from $t_0$ to $T = t_0 + 100M$. The maximum mismatch is taken at each $t_0$ and used as the numerical error at this point. This is the approach given in \cite{Finch:2021iip}. 

For the spatial mismatch plots in Fig. \ref{fig:Malpha}, the mapping is performed with the four simulation level and radius combinations. The spatial mismatch is calculated between the reconstruction from the preferred simulation and each of the remaining three, and the maximum value at each $t_0$ is used to quantify the numerical error. 

In Figs \ref{fig:reconstruction_t0}, \ref{fig:amplitude_ratios_t0}, \ref{fig:amplitude_ratios_t0_quad}, and \ref{fig:amplitude_ratios_t0_memory}, where a particular numerical value is recovered, a range is computed by determining the value for all simulation level and radius combinations and plotting the maximum and minimum at each $t_0$ as a translucent region around the preferred simulation value. 

\begin{figure}[ht]
    \centering
    \includegraphics[width=0.48\textwidth]{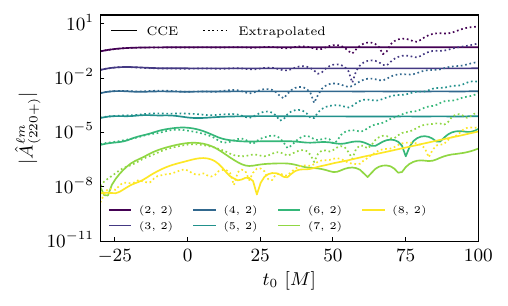}
    \caption{The magnitude of the decay-corrected amplitudes obtained from spatial mapping of the QNM $(2,2,0,+)$ are compared for two no-spin, equal-mass-ratio progenitor simulations. The bold lines are obtained using the CCE simulation 0001, and the dotted lines are obtained from simulation 0001 in the catalog of extrapolated waveforms.
    }
    \label{fig:sm_mismatch_loudest_ext_vs_CCE}
\end{figure}

For all analyses in this paper, CCE waveforms were used. To demonstrate the benefit of this method of extraction, Fig.~\ref{fig:sm_mismatch_loudest_ext_vs_CCE} compares the decay-corrected amplitudes obtained from spatial mapping of the $(2,2,0,+)$ QNM for two no-spin, equal-mass-ratio progenitor simulations. The first simulation, taken from the CCE catalog displays extremely stable amplitudes for $\ell \leq 4$ over a range of start times spanning at least 100$M$. This is in contrast to the extrapolated simulation, whose amplitudes appear to decay into noise significantly earlier, from around 25M. The difference in amplitude stability between simulations highlights the necessity of using the CCE waveforms, particularly at second order where modes are substantially quieter. 

\section{Computing quadratic structure predictions}

The mode-mixing coefficients ${}_{s}\mu^\beta_\alpha$ allow us to expand the spin-weighted spheroidal harmonics in the spin-weighted spherical harmonics basis,
\begin{align} \label{eq:Sexpansion}
    {}_{s}S_{\alpha}(\theta,\phi) = \sum_{\beta} {}_{s}\mu^{\beta}_{\alpha} \, {}_{s}Y^{\beta}(\theta,\phi),
\end{align}
where the mode-mixing coefficients are defined by the following integral:
\begin{align}
    {}_{s}\mu^{\beta}_{\alpha} = \int \dd\Omega\; {}_{s}S_{\alpha}(\theta, \phi) \left({}_{s}{Y}^\beta(\theta,\phi)\right)^*
    \label{mixing_calc}
\end{align}
This is consistent with the definition in Eq.~\ref{eq:mode_mixing}, but with the spin-weight dependence of the mode-mixing coefficients made explicit in the notation. Note that the mode-mixing coefficients are not spin-weight $s$ objects; here, the preceding subscript $s$ is intended as a reminder that these coefficients are to be used in the expansion of the spin-weight $s$ spheroidal harmonics. 
The mode-mixing coefficients are functions of the BH remnant spin $\chi_f$ (although this dependence is suppressed in our notation for clarity) and are computed using the \texttt{qnm} package.

In a similar way, the coefficients ${}_{ss'}\kappa^{\beta \beta' \beta''}$ allow us to re-expand products of two spin-weighted spherical harmonics in the spin-weighted spherical harmonics basis. Because the product of a spin-weight $s$ and a spin weight $s'$ field has spin-weight $s+s'$, this re-expansion is done in terms of the spherical harmonics with this spin weight, 
\begin{align} \label{eq:YYexpansion}
    {}_{s}Y^{\beta}(\theta,\phi) \, {}_{s'}Y^{\beta'}(\theta,\phi) = \sum_{\beta''} {}_{ss'}\kappa^{\beta \beta' \beta''} \; {}_{s+s'}Y^{\beta''}(\theta,\phi),
\end{align}
where the ${}_{ss'}\kappa^{\beta \beta' \beta''}$ coefficients are defined by the following integral,
\begin{align}
    {}_{ss'}\kappa^{\beta \beta' \beta''} = \int \dd\Omega\; {}_{s}Y_{\beta}(\theta, \phi) \, {}_{s'}Y_{\beta'}(\theta, \phi) \, \left({}_{s+s'}{Y}^{\beta''}(\theta,\phi)\right)^*.
\end{align}
These coefficients are not functions of the BH remnant spin and, like ${}_{s}\mu^\beta_\alpha$, they are not spin-weight fields.
They are computed using the \texttt{spherical} package.

These components allow us to expand the predictions for the quadratic shape functions $F_{\alpha \alpha'}$ discussed in the main text in the spin-weighted spherical harmonic basis

\textbf{Prediction A.} Substituting Eq.~\ref{eq:Sexpansion} twice into Eq.~\ref{eq:predA} and using Eq.~\ref{eq:YYexpansion} to re-expand the product of two spin-weight $s=-2$ spherical harmonics in terms of the spin-weight $s=-4$ spherical harmonics gives
\begin{align} \label{eq:predA_app}
    \mathbf{A:}\quad F_{\alpha \alpha'}(\theta,\phi) \propto \sum_{\beta} \sum_{\beta'} \sum_{\beta''} {}_{-2}\mu^\beta_{\alpha} \; {}_{-2}\mu^{\beta'}_{\alpha'} \; {}_{(-2)(-2)}\kappa^{\beta \beta' \beta''} {}_{-4}Y^{\beta''}(\theta,\phi).
\end{align}

\textbf{Prediction B.} Consider the B (i) prediction with $s=-2$ and $s'=0$. Substituting Eq.~\ref{eq:Sexpansion} twice into Eq.~\ref{eq:predBi} and using Eq.~\ref{eq:YYexpansion} to re-expand the product of a spin-weight $s=-2$ and a spin-weight $s=0$ spherical harmonic in terms of the spin-weight $s=-2$ spherical harmonics gives
\begin{align}
    \mathbf{B}\,\mathbf{(i):}\quad F_{\alpha \alpha'}(\theta,\phi) \propto \sum_{\beta} \sum_{\beta'} \sum_{\beta''} {}_{-2}\mu^{\beta}_{\alpha} \; {}_{0}\mu^{\beta'}_{\alpha'} \; {}_{(-2)(0)}\kappa^{\beta \beta' \beta''} {}_{-2}Y^{\beta''}(\theta,\phi).
\end{align}  
Similar expressions can be written for the other spin-weight combinations -- B (ii) and so on.

\textbf{Prediction C.} It is difficult to write this prediction concisely in the same notation used in this section. Instead, it is more convenient to compute the mixing directly from Eq. \ref{mixing_calc} using spheroidal harmonics with modified spheroidicity, determined using the \texttt{spheroidal} package.

\textbf{Prediction D.} The spin-weight raising operator satisfies $\eth {}_{s}Y^\beta(\theta,\phi) = \sqrt{(\beta_\ell-s)(\beta_\ell+s+1)}{}_{s+1}Y^\beta(\theta,\phi)$, where $\beta_\ell$ is the polar $\ell$ index from the $\beta$ tuple of spherical indices. From Eq.~\ref{eq:predD}, the term inside the parentheses is identical to prediction A and can be written as Eq.~\ref{eq:predA_app}, then acting twice with the spin-weight raising operator gives
\begin{align}
    \mathbf{D:}\quad F_{\alpha \alpha'}(\theta,\phi) \propto \sum_{\beta} \sum_{\beta'} \sum_{\beta''} {}_{-2}\mu^\beta_{\alpha} \; {}_{-2}\mu^{\beta'}_{\alpha'} \; {}_{(-2)(-2)}\kappa^{\beta \beta' \beta''} \; \sqrt{(\beta''_\ell+4)(\beta''_\ell-3)(\beta''_\ell+3)(\beta''_\ell-2)} \; {}_{-2}Y^{\beta''}(\theta,\phi).
\end{align}  

\end{document}